\documentclass[conference]{IEEEtran}
\usepackage{cite}
\usepackage[pdftex]{graphicx}
\DeclareGraphicsExtensions{.pdf,.jpeg,.png}
\usepackage{adjustbox}
\usepackage[cmex10]{amsmath}
\usepackage{eqnarray}
\usepackage{listings}
\usepackage{soul}
\usepackage{booktabs}
\usepackage{ctable}
\usepackage{multirow}
\usepackage{bigstrut}
\usepackage{flushend}

\usepackage{url}

\newcommand{\tool}{MARK}    

\newcommand{\code}[1]{{\footnotesize\textsf{#1}}}

\begin{document}

\title{Mining User Opinions in Mobile App Reviews:\\ A Keyword-based Approach}

\author{\IEEEauthorblockN{
Phong Minh Vu, Tam The Nguyen,
Hung Viet Pham,
Tung Thanh Nguyen}
\IEEEauthorblockA{Computer Science Department\\
Utah State University\\
\{phong.vu,tam.nguyen,hung.pham\}@aggiemail.usu.edu\\
tung.nguyen@usu.edu}}
\maketitle

\begin{abstract}
User reviews of mobile apps often contain complaints or suggestions which are valuable for app developers to improve user experience and satisfaction. However, due to the large volume and noisy-nature of those reviews, manually analyzing them for useful opinions is inherently challenging. To address this problem, we propose {\tool}, a keyword-based framework for semi-automated review analysis. {\tool} allows an analyst describing his interests in one or some mobile apps by a set of keywords. It then finds and lists the reviews most relevant to those keywords for further analysis. It can also draw the trends over time of those keywords and detect their sudden changes, which might indicate the occurrences of serious issues. To help analysts describe their interests more effectively, {\tool} can automatically extract keywords from raw reviews and rank them by their associations with negative reviews. In addition, based on a vector-based semantic representation of keywords, {\tool} can  divide a large set of keywords into more cohesive subsets, or suggest keywords similar to the selected ones.
\end{abstract}


%
\IEEEpeerreviewmaketitle

\section{Introduction}
Mobile apps are the software applications developed specially for mobile devices such as smartphones and tablets. As the use of mobile devices explodes, developing mobile apps becomes a popular and profitable business in software development. However, it is also a highly competitive business, as millions and counting apps of different categories are made available on app markets. Since the revenue and profit of a mobile app is often proportional to the size of its userbase, improving user experience and satisfaction to retain existing users and attract new ones is of important to its developers. User opinions like complaints or suggestions would be valuable for that task.

As mobile app markets typically provide rating and reviewing mechanisms, reviews from users of apps purchased on those markets provide an important source of user opinions. However, analyzing those reviews manually for useful opinions would be inherently challenging. First, a popular app with millions of users often gets thousands of reviews each day and reading all of those reviews would be very time-consuming. In addition, user reviews of mobile apps are often noisy. They can have typos, acronyms, abbreviations, emoji icons, etc. Even worse, prior research reports that more than 60\% of user reviews do not contain useful opinions~\cite{ARminer}.

In this paper, we introduce {\tool} (\emph{\underline{M}ining and \underline{A}nalyzing \underline{R}eviews by \underline{K}eywords}), a semi-automated framework for mining user opinions from user reviews of mobile apps. Considering this mining task as an information retrieval problem, {\tool} follows a keyword-based approach. That is, it allows a review analyst to specify his/her interests in one or some mobile apps by a set of keywords. Then, it uses those keywords to search for and visualize the most relevant reviews, expecting them to contain opinions usefully matched the analyst's specified interests. The key departure of {\tool} from a typical information retrieval system is that it employs several automated, customized techniques for extracting keywords from raw reviews, ranking those keywords based on review ratings and occurrence frequencies, grouping related keywords, searching for reviews that are relevant to a set of keywords, visualizing their occurrences over time, and reporting if such occurrences contain unusual patterns.

Let us illustrate {\tool} and those techniques via an example. Assume that a review analyst is interested in negative user opinions about Facebook Messenger, one of the most popular mobile apps with around 700 millions active users by July 2015~\cite{facebookStatistic}. Initially, he has no idea about which aspects of the app that get negative opinions. Thus, {\tool} uses its customized keyword extraction technique discussed in details in Section~\ref{sec:normalize} to extract all potential keywords from raw reviews of this app. Then, it ranks those keywords based on a negative scoring scheme discussed in details in Section~\ref{sec:preprocessing} and presents the ranked list to the analyst. Table~\ref{tab:example.keyword} illustrates the top ones among them.
\begin{table}[t]
  \caption{Negative keywords for Facebook Messenger}
  \centering
  \sf
    \begin{adjustbox}{max width=0.47\textwidth}

      	  \renewcommand{\arraystretch}{0.6}
    \begin{tabular}{rlrlrlrl}
    \toprule
    \textbf{Rank} & \textbf{Keyword} & \textbf{Rank} & \textbf{Keyword} & \textbf{Rank} & \textbf{Keyword} & \textbf{Rank} & \textbf{Keyword}\\
    \midrule
    1 & battery & 6 & expire & 11 & phone & 16 & say\\
    2 & message & 7 & drain & 12 & app & 17 & space\\
    3 & download & 8 & crash & 13 & keep & 18 & use\\
    4 & install & 9 & fix & 14 & facebook & 19 & freeze\\
    5 & session & 10 & log & 15 & reinstall & 20 & network\\
    ... & ... & ... & ... & ... & ... & ... & ...\\
    \bottomrule
    \end{tabular}%
     \end{adjustbox}
  \label{tab:example.keyword}%
\end{table}%

\begin{table}[t]

  \caption{Clusters of negative keywords}
  \centering
  \sf
    \begin{tabular}{lll}
    \toprule
    \textbf{Energy consumption} & \textbf{Unrecoverable error} & \textbf{Authentication}\\
    \midrule
    battery, drain, & crash, freeze, & session, login,\\
    hog, consume & hang, break & fail, connect\\
    \bottomrule
    \end{tabular}%
  \label{tab:example.cluster}%
\end{table}%

\begin{table}[t]
  \caption{Expanding keywords for ``battery'' and ``drain''}
  \centering
  \sf
    \begin{tabular}{c}
\toprule
heat, hog, usage, consumption, consume,\\
battery, drain, hogger, overheat, eater,\\
eat, drainer, power, ram, cpu, storage, memory\\
\bottomrule
\end{tabular}%
  \label{tab:example.expand}%
\end{table}%

We could see in the table that keywords are often related and indicate more general concerns. For example both keywords \code{``crash''} and \code{``freeze''} could be used to describe the app's status when an \emph{``unrecoverable error''} occurs. Or, \code{``battery''} and \code{``drain''} often go together to describe the bad \emph{``energy consumption''} of the app. Therefore, {\tool} can cluster, i.e. divide the listed keywords into smaller subsets, each for a more general concern. Table~\ref{tab:example.cluster} illustrates the resulted clusters produced for Facebook Messenger.

This clustering task is based on Word2Vec, a distributed, vector-based representation of words~\cite{word2vec}. Word2Vecvec represents each word in the vocabulary as a high dimensional vector and learns those vectors from a large corpus of text such that words having similar or related syntactic roles or semantic meanings would have similar vectors. Based on Word2Vec, {\tool} can divide a set of keywords into smaller subsets of related ones by applying $K$-mean\cite{kmean}, a similarity-based clustering algorithm on the vectors representing those keywords. Details about Word2Vec and {\tool}'s clustering technique will be discussed in Section \ref{sec:dividing}.

Instead of dividing a large set of keywords into smaller subsets, {\tool} can expand a small set of keywords into a bigger one, also based on the vector-based similarity of the keywords. This case often happens when the analyst has some ideas about the opinions he is looking for and wants to explore them in a broader context. For example, when analyzing the keywords in Table~\ref{tab:example.keyword}, the analyst sees the keyword \code{battery} and is interested in the \emph{``energy consumption''} aspect of this app. He assumes keywords \code{``battery''} and \code{``drain''} to be related to this topic, but expects users using other keywords to describe it. Therefore, he requests {\tool} to expand his initial keyword set \{\code{``battery''}, \code{``drain''}\}.
Table~\ref{tab:example.expand} shows the expanding results, with newly discovered keywords like \code{``heat''}, \code{``power''}, and \code{``usage''}.

Once the analyst specifies a set of keywords (via clustering or expanding) that matches his interests, {\tool} can query its review database and returns ones that are most relevant to that keyword set. This querying task is based on the standard Vector Space Model~\cite{word2vec}. That is, {\tool} applies the \emph{tf.idf}\cite{manning2008introductionInfoRetrieval} weighting scheme on the keywords and measures the relevance between the given keyword set to a review based on the cosine similarity of their \emph{tf.idf} feature vectors. Table~\ref{tab:example.query} illustrates some reviews queried for the keyword set in Table~\ref{tab:example.expand}. As seen, those reviews contain several (negative) user opinions about the \emph{``energy consumption''} aspect of this app.

\begin{table}[t]
  \caption{Review search result for keywords related to ``energy consumption''}
  \centering
  \sf
    \begin{adjustbox}{max width=0.48\textwidth}
    \begin{tabular}{l}
    \toprule
\parbox{\linewidth}{\textbf{Battery} \textbf{drain}. Latest version usually destroying my \textbf{battery} life, \textbf{consuming} almost a third of my phones \textbf{power} \textbf{consumption} and I haven't opened the app or gotten a single message all day!!! Please fix!}\\
\midrule
\parbox{\linewidth}{Batt \textbf{hogger}. Disabled notification sounds \& vibrate. Still \textbf{drains} more \textbf{battery} than the screen itself. Can do without.}\\
\midrule
\parbox{\linewidth}{Excessive \textbf{battery} \textbf{usage}, \textbf{overheating}. Excessive \textbf{CPU} and \textbf{battery} \textbf{usage} is leading to quickly \textbf{drained} \textbf{battery} and \textbf{overheating} even when my tablet is sleeping. Uninstalling until this is fixed; it's killing my \textbf{battery}.}\\
    \bottomrule
    \end{tabular}%

      \end{adjustbox}
  \label{tab:example.query}%
\end{table}%

{\tool} can also visualize and analyze the occurrences of a keyword set overtime for abnormal patterns. Prior studies~\cite{ARminer, Wiscom} suggest that when a new version of an app is released with some severe defects or issues making many users unsatisfied, there are often sudden changes in occurrences of related topics in user reviews. For example, in Feb 2015 a new version of Facebook Messenger is released with a critical error in the syncing functionality~\cite{facebook} which generates extremely high CPU and network activities, and thus causes severe battery draining. This leads to an unusual occurrence of user reviews discussing \emph{``energy consumption''} topic and related keywords. To detect such abnormalities, {\tool} considers the keywords' occurrence counts as a time series, computes its simple moving average (SMA) and the differences between actual counts and its SMA values. If a difference value is significant higher than the standard deviation of SMA values, it is considered as a sudden change in the corresponding occurrence count. Figure~\ref{fig:trendeval} illustrates the analysis result of {\tool} on this example. Section IV-B will discuss the technique in more details.

\begin{figure}[t]
\centering
\includegraphics[width=0.4\paperwidth]{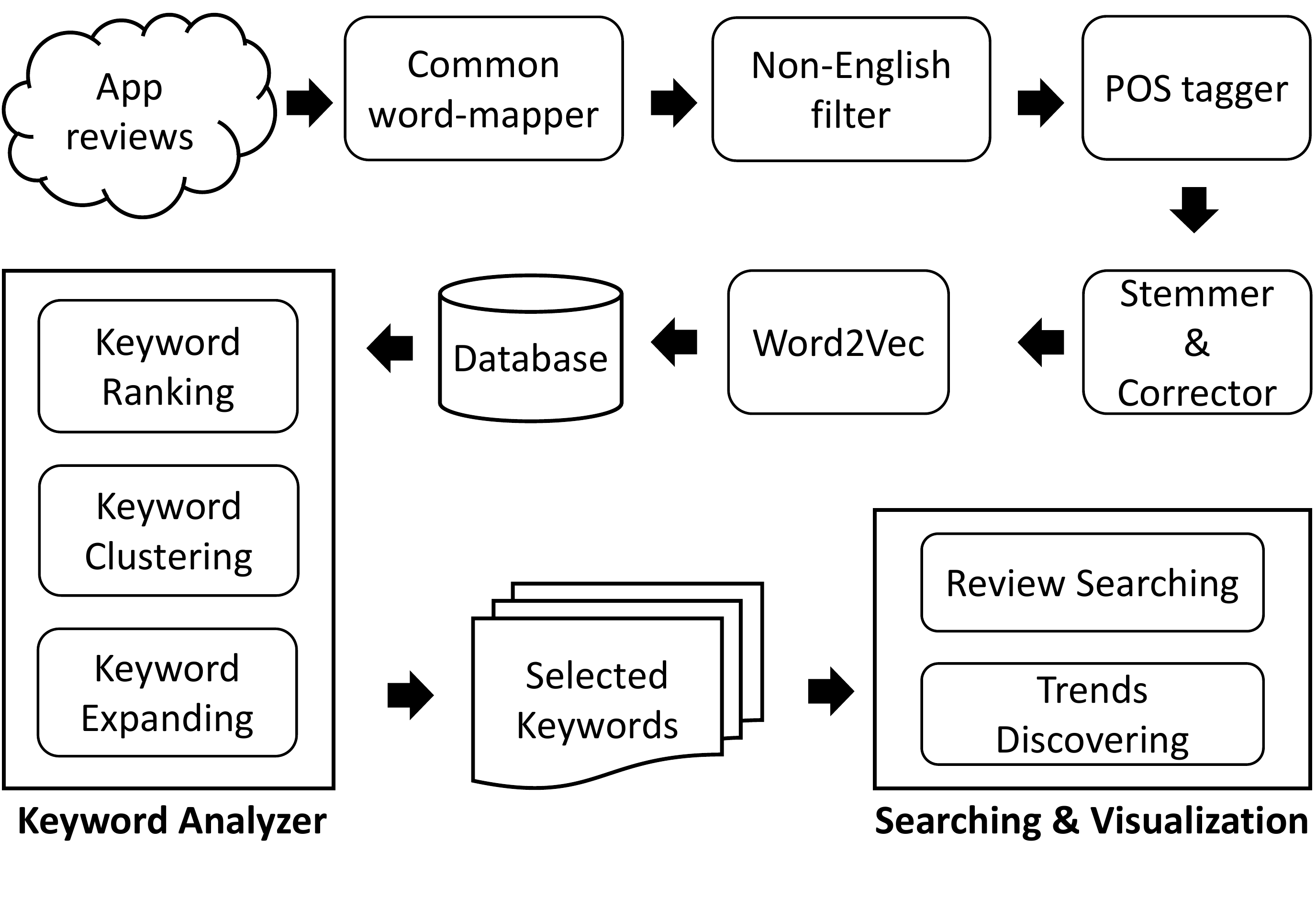}
\caption{System Architecture and Processing Pipeline of {\tool}}
\label{fig:overview}
\end{figure}

We have developed and deployed {\tool} as a web-based tool available online at \url{http://useal.cs.usu.edu/mark}. 
Figure~\ref{fig:overview} summarizes its system architecture and processing pipeline. In general, {\tool} works in three stages. In the pre-processing stage, app reviews are crawled from designated online app stores such as Google Play or Apple App Store and are tokenized as sequences of words. Then, the Common Word Mapper replaces frequently misspelled words and popular acronyms and abbreviations with their corrected forms. The Non-English Filter will analyze the reviews and discard ones not written in English. In the next step, word sequences of unfiltered reviews are analyzed by a part-of-speech (PoS) tagger. After this step, {\tool} keeps only nouns and verbs and transforms them to their base forms via stemming and spell-correction. After that, Word2Vec component processes those words to produce their semantic vectors. In the two remaining stages, {\tool} allows users select keywords of their interests by using Keyword Ranking, Keyword Clustering, and Keywords Expanding components. Once the keywords are selected, the user can view the related reviews provided by the Research Searching component or view the occurrence trends of those keywords via the Trends Discovering component.



The remaining of this paper is organized as the following.
Section \ref{sec:DataCollectionPreprocessing} describes {\tool}'s keyword extraction and its customized regulation algorithms. Section \ref{sec:keyword} discusses keyword analysis techniques used in {\tool}. Section \ref{sec:trend} presents our trend abnormalities detection and visualization technique. Section \ref{sec:eval} presents our empirical evaluations and case studies. Related works are discussed in Section \ref{sec:relatedwork} and we conclude our study in Section \ref{sec:conclusion}.

\section{Keyword Extraction}
\label{sec:DataCollectionPreprocessing}
In this section, we describe how {\tool} pre-processes raw review data and extracts keywords from them. We first discuss the common issues of processing mobile app reviews including misspelled words and typos, acronyms, abbreviations, slangs, non-English text, etc. After that, we propose a keyword extraction procedure with several customized algorithms to address such problems.



%
%

\subsection{Reviews Pre-processing}\label{sec:preprocessing}

\subsubsection{Misspelled words, acronyms, and abbreviations}
\label{sec:misspell}
Mobile app users are most likely to write their reviews on mobile devices. However, most of those devices have small sizes and no physical keyboards, making it difficult and time-consuming for typing. This situation leads to the frequent occurrences of typos (i.e. misspelled words), acronyms, and abbreviations in user reviews of those mobile apps.

\begin{table}
  \centering
  \sf
  \caption{Top 10 misspelled words collected from 300,000 reviews}
  \begin{adjustbox}{max width=0.5\textwidth}

   \renewcommand{\arraystretch}{0.6}
    \begin{tabular}{llr|llr}
    \toprule
    \textbf{Misspelled} & \textbf{Corrected} & \textbf{Count} & \textbf{Misspelled} & \textbf{Corrected} & \textbf{Count}\\
    \midrule
    usefull & useful & 1104 & appp  & app & 351 \\
    excelent & excellent & 452 & watsapp & whatsapp & 348 \\
    helpfull & helpful & 417 & coool & cool & 261 \\
    dosent & doesn't & 374 & exelent & excellent & 241 \\
    frnds & friends & 352 & aplication & application & 228\\
    \bottomrule
    \end{tabular}%
  \end{adjustbox}
  \label{tab:misspelled}%
\end{table}%

We performed a preliminary analysis of 300,000 reviews collected from Google Play during January 2015. Checking against an English dictionary of more than 150,000 common words, we found nearly 5,500 words do not belong to that dictionary and occur at least 20 times in those reviews. Table \ref{tab:misspelled} shows the top 10 among them. As illustrated in the table, most are misspelled words like \code{``usefull''} or \code{``excelent''}. Some are abbreviations (e.g.  \code{``frnds''} is an abbreviation of \code{``friends''}) and some refer to special names (e.g. \code{``watsapp''} is a misspelled word for \code{``WhatsApp''} - a popular messaging app for mobile devices).

Common spell correctors like FAROO~\cite{faroo} or Peter Norvig's corrector~\cite{norvicspell} would not work well for such abbreviations and special names. Therefore, we decided to use a custom dictionary for the most frequent out-of-dictionary words we detected in the aforementioned preliminary analysis. To develop this dictionary, our team of four researchers analyzed each word and mapped it to the most plausibly corrected form. If a word is not so obvious, we will read all reviews containing that word to infer its corrected form from the context. For example, we consider \code{intagram} as a misspelled form of \code{Instagram}, a popular image-sharing service. Our dictionary also contains app names, acronyms, abbreviations, and technical terms that would not appear in common English dictionaries, like \code{``reinstall''}, \code{``hashtag''}, or \code{``xml''}. This dictionary is made available on {\tool}'s website.

\subsubsection{Non-English reviews}
Mobile apps are available globally, their users can come from different countries and cultures and thus, might write reviews in languages different from English. However, we only	 focus on analyzing English reviews (which are the most popular), non-English reviews would introduce noises to the analyses and thus, should be filtered. Prior work~\cite{Wiscom} filters non-English reviews based on the occurrences of non-ASCII characters. However, many languages can be written with ASCII characters, e.g. Spanish, French, or even Vietnamese. Therefore, we decide to develop a custom filtering algorithm for non-English reviews.

We performed a preliminary analysis on non-English reviews for some insights on their characteristics. We randomly selected 400 reviews and manually labeled them as English (245 reviews) or non-English (155 reviews). We observed with no surprise that most words in non-English reviews do not appear in our English dictionary. However, sometimes a non-English review can contain English words (e.g. a technical term like \code{``email''}) or words in foreign languages that accidentally have the same spellings with English words (e.g. \code{``can''} is a Vietnamese word that is also a meaningful English word). This is also true for English reviews, i.e. sometimes they also contain words not appear in English dictionaries, e.g. misspelled words that have not been corrected in previous pre-processing steps.

Designed based on such observations, our filtering algorithm considers a review to be written in English if the ratio of its single words (i.e. unigram) appearing in our English dictionary (which also includes special names, acronyms, abbreviations, and technical terms found in our previous analysis) exceeds a preset threshold. To improve the accuracy of this algorithm, we also compute the ratio of pairs of consecutive English words (bigram) in each review and compare it to another preset threshold. This rule is based on the assumption that English speakers would not make too many mistakes repeatedly (e.g. one can misspell a word but would be less likely to misspell two words in a row).

\subsubsection{Word stemming and Part-of-Speech tagging}
\label{sec:normalize}

\begin{figure}[t]
\begin{lstlisting}[language=Pascal, basicstyle=\scriptsize\ttfamily, numbers=right, numbersep=-5pt,mathescape]
function Stem(a review $r$)
  for each sentence $c \in r$
    for each verb/noun $w \in$ PoS($c$)
      if $w$ is an irregular verb
        @\textbf{return}@ IrregularVerbTable[$w$]
      else
        apply stemming rules on $w$
        @\textbf{return}@ SpellingCorrect($w$)
\end{lstlisting}
\caption{Customized Stemming Algorithm}
\label{fig:stem.algo}
\end{figure}

Information retrieval system commonly employ stemming, i.e. reducing words to its simplest form, to improve their searching capacity. However, common stemmers like Porter's Snowball~\cite{porter2001snowball} often over-stem words. For example, it would stem the words \code{``conspiracy''}, \code{``conspirator''}, \code{``conspire''}, \code{``conspired''} and \code{``conspiring''} as \code{``conspir''}. This is not desirable for our keyword-based analysis because we want the keywords to retain their original meanings. Lemmatization tools like Stanford Lemmatizer~\cite{manning:SFlemma} do not over-stem words, but might be too slow for our review data.

In addition, we expect review analysts to be interested in user opinions related to features or issues of mobile apps and such features and issues would be described by verbs (e.g. \code{``freeze''} or \code{``crash''}) and nouns (e.g. \code{``battery''} or \code{``screen''}). Adjectives and adverbs often describe additional information to the main nouns and verbs (e.g. \code{``slow''}, \code{``bad''}, or \code{``great''}), thus do not provide references to functions or features of mobile apps. Other parts of speech like pronouns or prepositions even contribute less to the general understanding of user opinions (they are often called stop words and removed by information retrieval systems). Therefore, we need to identify the part-of-speech (PoS) tags of the words and keep only verbs and nouns as potentially useful keywords.

Due to such issues, we decide to design a customized stemmer to work exclusively for verbs and nouns. The main purpose of this stemmer is to reduce verbs and nouns in different tenses and forms back to their base forms, e.g. \code{``frozen''} to \code{``freeze''} or \code{``notifications''} to \code{``notification''}. Figure~\ref{fig:stem.algo} summarizes our stemming algorithm. The stemmer first breaks the given review into sentences and employs the Stanford PoS tagger~\cite{SFpostagger} to find verbs and nouns in each sentence. If a verb is an irregular verb, its base form will be retrieved directly from a pre-defined table of irregular verbs~\cite{IrregularEnglishverbs}. Otherwise, stemming rules in Table~\ref{tab:stemrules} will be applied directly on the word.

\begin{table}[t]
  \centering\sf
  \caption{Stemming Rules}
  \begin{adjustbox}{max width=0.55\textwidth}
      \setlength{\tabcolsep}{2pt}
      \tiny
      \renewcommand{\arraystretch}{0.6}
    \begin{tabular}{|l|l|l|l|l|}
    \hline
    \textbf{Case} & \textbf{Pre-condition} & \textbf{Condition} & \textbf{Action} & \textbf{Note/ Example} \bigstrut\\
    \hline
    \multirow{10}[20]{*}{\parbox[t]{.1\linewidth}{plural noun OR verb- present tense- 3rd person singular}} & \parbox[t]{.15\linewidth}{length $<$ 4 OR not end with 's'} &       & \parbox[t]{.15\linewidth}{do nothing} & \parbox[t]{.2\linewidth}{Fail safe if PoS tagger makes mistake} \bigstrut\\
\cline{2-5}          & \multirow{8}[16]{*}{\parbox[t]{.15\linewidth}{end with 'es', X-Y-Z: last 3 characters before 'es'}} & \parbox[t]{.2\linewidth}{YZ is a SPECIAL PAIR} & \parbox[t]{.15\linewidth}{remove 's'} & \parbox[t]{.2\linewidth}{divestitures $\rightarrow$ divestiture} \bigstrut\\
\cline{3-5}          &       & \parbox[t]{.2\linewidth}{Z $=$ 'i'} & \parbox[t]{.15\linewidth}{remove 'es' and change 'i' to 'y' if Y is a vowel, just remove 'es' otherwise.} & \parbox[t]{.2\linewidth}{studies $\rightarrow$ study} \bigstrut\\
\cline{3-5}          &       & \parbox[t]{.2\linewidth}{YZ $=$ 'ss'} & \parbox[t]{.15\linewidth}{remove 'es'} & \parbox[t]{.2\linewidth}{masses $\rightarrow$ mass} \bigstrut\\
\cline{3-5}          &       & \parbox[t]{.2\linewidth}{Z $=$ 's'} & \parbox[t]{.15\linewidth}{remove 's'} & \parbox[t]{.2\linewidth}{bases $\rightarrow$ base} \bigstrut\\
\cline{3-5}          &       & \parbox[t]{.2\linewidth}{Z $=$ 'x', 'o', 'z'} & \parbox[t]{.15\linewidth}{remove 'es'} & \parbox[t]{.2\linewidth}{foxes $\rightarrow$ fox} \bigstrut\\
\cline{3-5}          &       & \parbox[t]{.2\linewidth}{YZ $=$ 'ch', 'sh'} & \parbox[t]{.15\linewidth}{remove 'es'} & \parbox[t]{.23\linewidth}{approaches $\rightarrow$ approach} \bigstrut\\
\cline{3-5}          &       & \parbox[t]{.2\linewidth}{Z $=$ 'h'} & \parbox[t]{.15\linewidth}{remove 's'} & \parbox[t]{.2\linewidth}{breathes $\rightarrow$ breathe} \bigstrut\\
\cline{3-5}          &       & \parbox[t]{.2\linewidth}{everything else} & \parbox[t]{.15\linewidth}{remove 's'} & \parbox[t]{.2\linewidth}{combines $\rightarrow$ combine} \bigstrut\\
\cline{2-5}          & \parbox[t]{.15\linewidth}{end with 's'} &       & \parbox[t]{.15\linewidth}{remove 's'} & \parbox[t]{.2\linewidth}{annoys $\rightarrow$ annoy} \bigstrut\\
    \hline
    \multirow{7}[14]{*}{\parbox[t]{.1\linewidth}{verb-past tense OR verb-past participle}} & \parbox[t]{.15\linewidth}{length $<$ 5 OR not end with 'ed'} &       & \parbox[t]{.15\linewidth}{do nothing} & \multirow{2}[4]{*}{\parbox[t]{.2\linewidth}{Fail safe if PoS tagger makes mistake}} \bigstrut\\
\cline{2-4}          & vowel count $=$ 1 &       & do nothing &  \bigstrut\\
\cline{2-5}          & \multirow{4}[8]{*}{\parbox[t]{.15\linewidth}{X-Y-Z: last~3 characters~before 'ed'}} & \parbox[t]{.2\linewidth}{length $=$ 5 \&\& Y is a vowel} & \parbox[t]{.15\linewidth}{remove 'd'} & \parbox[t]{.2\linewidth}{fired $\rightarrow$ fire} \bigstrut\\
\cline{3-5}          &       & \parbox[t]{.2\linewidth}{Z $=$ 'i'} & \parbox[t]{.15\linewidth}{remove 'ed', change 'i' to 'y'} & \parbox[t]{.2\linewidth}{implied $\rightarrow$ imply} \bigstrut\\
\cline{3-5}          &       & \parbox[t]{.2\linewidth}{vowel count $=$ 2 \&\& YZ $=$ 'tt', 'nn', 'rr', 'dd', 'mm', 'ff', 'gg', 'pp', 'bb'} & \parbox[t]{.15\linewidth}{remove 'ed' and one last consonant} & \parbox[t]{.2\linewidth}{hogged $\rightarrow$ hog} \bigstrut\\
\cline{3-5}          &       & \parbox[t]{.2\linewidth}{P(e$|$XYZ) * P(' '$|$YZe) $>$ P(' '$|$XYZ)} & \parbox[t]{.15\linewidth}{remove 'd'} & \parbox[t]{.2\linewidth}{3-gram model} \bigstrut\\
\cline{2-5}          & \parbox[t]{.15\linewidth}{everything else} &       & \parbox[t]{.15\linewidth}{remove 'ed'} & \parbox[t]{.2\linewidth}{acted $\rightarrow$ act} \bigstrut\\
    \hline
    \multirow{6}[12]{*}{\parbox[t]{.1\linewidth}{verb-present participle (gerund)}} & \parbox[t]{.15\linewidth}{length $<$ 6 OR not end with 'ing'} &       & \parbox[t]{.15\linewidth}{do nothing} & \multirow{2}[4]{*}{\parbox[t]{.2\linewidth}{Fail safe if PoS tagger makes mistake}} \bigstrut\\
\cline{2-4}          & vowel count $=$ 1 &       & do nothing &  \bigstrut\\
\cline{2-5}          & \multirow{3}[6]{*}{\parbox[t]{.15\linewidth}{X-Y-Z: last~3 characters~before 'ing'}} & \parbox[t]{.2\linewidth}{length $=$ 6 \&\& Y is a vowel} & \parbox[t]{.15\linewidth}{remove 'ing', append 'e'} & \parbox[t]{.2\linewidth}{firing $\rightarrow$ fire} \bigstrut\\
\cline{3-5}          &       & \parbox[t]{.2\linewidth}{vowel count $=$ 2 \&\& YZ $=$ 'tt', 'nn', 'rr', 'dd', 'mm', 'ff', 'gg', 'pp', 'bb'} & \parbox[t]{.15\linewidth}{removing 'ing' and one last consonent} & \parbox[t]{.2\linewidth}{hogging $\rightarrow$ hog} \bigstrut\\
\cline{3-5}          &       & \parbox[t]{.2\linewidth}{P(e$|$XYZ) * P(' '$|$YZe) $>$ P(' '$|$XYZ)} & \parbox[t]{.15\linewidth}{remove 'ing', append 'e'} & \parbox[t]{.2\linewidth}{3-gram model} \bigstrut\\
\cline{2-5}          & \parbox[t]{.15\linewidth}{everything else} &       & \parbox[t]{.15\linewidth}{remove 'ing'} & \parbox[t]{.2\linewidth}{acting $\rightarrow$ act} \bigstrut\\
    \hline
    \end{tabular}%
  \end{adjustbox}
  \label{tab:stemrules}%
\end{table}%

Those rules are designed based on English grammar and some rules of the Snowball stemmer. As seen in the table, we only stem plural nouns and verbs in the present tense having an \code{``s''} or \code{``es''} suffix, verbs in the past or past perfect tenses having an \code{``ed''} suffix, and verbs in the present continuous tense having an \code{``ing''} suffix. The stemming rules aim to remove those suffixes and convert the noun or verb back to their original forms before those suffixes are added.

Although less likely, our stemmer might still over-stem words. To address this problem, we train a tri-gram language model at character level on a dataset of one billion characters containing text crawled from Wikipedia~\cite{wiki1bil}, and use this model to check if a word is over-stemmed and fix it. For example, word \code{``analyzing''} is a verb in the present continuous tense. However, when our stemmer removes its suffix \code{``ing''}, its new form \code{``analyz''} is over-stemmed because our tri-gram model suggests that \code{``lyz''} is less likely to appear as ending characters of an English word than \code{``yze''} and \code{``e''} is more likely to appear after \code{``lyz''}. Thus, character \code{``e''} is added to the word, producing \code{``analyze''} as the final (stemmed) form of \code{``analyzing''}.

We further fix the over-stemming problem by running the FAROO Spelling Corrector after applying our stemming rules in Table~\ref{tab:stemrules}. We choose this spelling corrector because it is very time-efficient. We also train it with the aforementioned one-billion-character dataset. It should be noted that stemming and spell correcting are applied only for common words and not for words in the dictionary of popular misspelled words and special names discussed in Section~\ref{sec:misspell} because the corrections for them have been included in that dictionary.

\section{Keyword Recommendation}
\label{sec:keyword}
{\tool} can extract a vast number of keywords from a large amount of user reviews. Therefore, it has three recommendation techniques to help analysts navigate those keywords and describe their interests more accurately and effectively. The first is a ranking technique to identify keywords that often associate with negative reviews, which are likely to describe the issues or features of mobile apps that make users unhappy. Two other techniques are clustering and expanding, which could help an analyst divide a large set of keywords into smaller, more cohesive subsets or expand a small set of keywords into a larger, more comprehensive one. This section will describe those techniques in details.

\subsection{Keyword Ranking}	

Information retrieval systems often rank their results to help their users focus on the most relevant ones. {\tool} also ranks the keywords it extracts from user reviews to help review analysts select the most important keywords. This ranking technique is designed based on the assumption that keywords frequently appearing with negative reviews are likely to describe the issues or features of the apps that cause bad user experience, i.e. making users unhappy or unsatisfied. Thus, those keywords would be of interested to app developers and review analysts because they can help to identify the bad aspects of an app and the user opinions about such aspects.

\begin{figure}[t]
\centering
\includegraphics[width=0.4\textwidth]{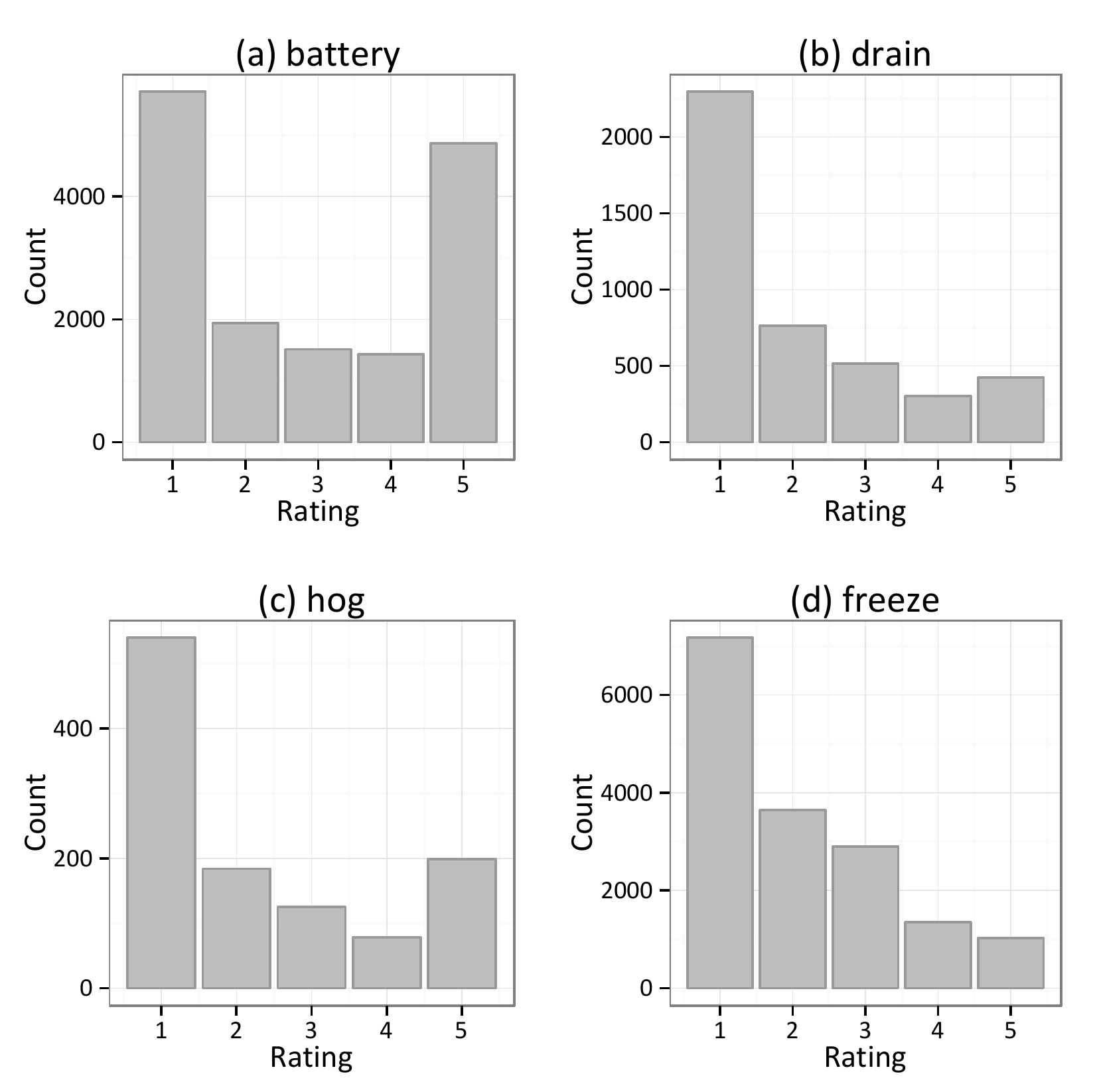}
\caption{Distribution of ratings for some keywords}
\label{fig:keywords}
\end{figure}

Table~\ref{tab:example.query} and Figure~\ref{fig:keywords} provide some anecdotal evidence of that assumption. The reviews in the table show that users use words like \code{``battery''}, \code{``drain''}, and \code{``hog''} when complaining about the excessive energy consumption of some mobile apps. Figure~\ref{fig:keywords} shows the numbers of reviews containing those words grouped by their ratings in {\tool} most recent review dataset of more than two millions reviews. As seen, those words appears much more in negative (i.e. 1- or 2-star) reviews than in positive (i.e. 4- or 5-star) reviews. For example, \code{``drain''} appears in {3,060} negative reviews and {726} positive reviews, which is four times more negative than positive. \code{``battery''} is not as excessive, but the number of negative reviews is still higher than that of positive reviews ({7,645 vs 6,297}). The figure also shows the counts for word \code{``freeze''} which is often used to describe or  complain about apps that suddenly stop working or responding. As seen \code{``freeze''} appears in negative reviews five times more than in positive reviews ({10,814 vs 2,378}).

Based on that assumption, {\tool} ranks the keywords by their association with negative reviews. It measures such association using a metric that we call \emph{contrast score}: $$\varphi = \frac {n}{p}\times (n-p)$$ In this formula, $n$ and $p$ are correspondingly the counts of negative (i.e. having ratings of 1 and 2 stars) and positive (i.e. having ratings of 4 and 5 stars) reviews containing the keyword of interest. For example, as shown in Figure~\ref{fig:keywords}, keyword \code{``freeze''} has $n = {10,814}$ negative reviews and $p = {2,378}$ positive reviews. Thus, its contrast score is ${{10,814}}/{2,378} \times ({10,814} - {2,378}) = {38,362.87}$. The contrast scores of \code{``drain''} and \code{``battery''} are ${9,837.52}$ and ${1,636.57}$, respectively.

In addition to the ratio of negative and positive reviews, the formula also combines their absolute difference because that will give more score to keywords occurring frequently in user reviews. Such a keyword is more likely to describe an issue that affect many users. This also help to downgrade keywords with low frequencies but extremely high contrast ratio, e.g. $n = 10$ and $p = 1$.

We also consider two other scoring techniques: the Pearson (linear) \emph{correlation} of the ratings and their counts, and the \emph{skewness} of the distribution of the ratings. The relevance of those scores can be inferred from Figure~\ref{fig:keywords}. For example, for keyword \code{``freeze''}, higher ratings (e.g. 4 or 5 stars) have lower counts, leading to a negative linear correlation of {-0.937}. For the same reason, the distribution of the ratings is highly \emph{left-skew}, with the skewness of {8.719}.

We performed a preliminary analysis to compare those three scoring techniques. It is surprised that they produced nearly identical results. For example, the sets of top 1,000 keywords ranked by those three metrics share 96\% to 99\% of their keywords. Due to this preliminary result, we use our contrast score as the main ranking technique in the final version of {\tool} because it can be computed very efficiently and incrementally (e.g. we just need the counts for computation and we could store and update the counts when new data are crawled and pre-processed).

\subsection{Keyword Clustering and Expanding}
\label{sec:dividing}
Because {\tool} performs review search and trend visualization based on keywords provided by the analysts, the more relevant of those keywords to the issues of interest, the more accurate and effective the search and visualization will be. In practice, users often use similar or related keywords to describe an issue or a feature. For example, we have seen in Table~\ref{tab:example.query} that \code{``battery''}, \code{``power''}, \code{``drain''}, and \code{``hog''} are used together to describe the issue of excessive energy consumption. Therefore, grouping similar or related keywords would help analysts obtain more better analysis results.

{\tool} provides two different grouping functions for its keywords. It can divide (i.e. cluster) a large set of keywords into into smaller, more cohesive subsets. For example, an analyst can start by selecting the top 100 keywords from reviews of an app and then cluster them into 10 subsets, expecting each set to describe a specific issue of the app. In contrast, when the analyst starts with just a few keywords for a specific concern of her interests, {\tool} can expand that set into a larger, more comprehensive one by adding many more keywords that are similar or related to the selected ones.

Both of these grouping techniques require a similarity measure for keywords. Traditional approaches often use WordNet~\cite{miller1998wordnet} to obtain similar words (i.e. synonyms). However, {\tool} utilizes Word2Vec, a distributed, data-driven, vector-based representation of words~\cite{word2vec}. This technique represents each keyword as a high dimensional vector which it learns from our review data. Because its learning algorithm produces similar vectors for keywords having similar or related syntactic roles or semantic meanings, {\tool} measures the similarity of keywords based on the distances of their vectors.

\begin{figure}[t]
\begin{lstlisting}[language=Pascal, basicstyle=\scriptsize\ttfamily, numbers=right, numbersep=-5pt,mathescape]
function Cluster(keywordset $S$, number of clusters $k$)
  for $i$ = 1 to $k$
    centroid[$i$] = a random vector $\in S$
  repeat
    for each keyword $v \in S$
      cluster[$v$] = $i$ where $i$ is the centroid closest to $v$
    for $i$ = 1 to $k$
      centroid[$i$] = meanvector($\forall v$ where cluster[$v$] = $i$)
  until cluster has no change
  @\textbf{return}@ cluster
\end{lstlisting}
\caption{Keyword Clustering Algorithm ($k$-mean)}
\label{algo:cluster}
\end{figure}

\begin{figure}[t]
\begin{lstlisting}[language=Pascal, basicstyle=\scriptsize\ttfamily, numbers=right, numbersep=-5pt,mathescape]
function Expand(keywordset $S$, vocabulary $V$, threshold $\delta$)
  repeat
    for each keyword $v$ $\in$ $V-S$
      $c$ = meanvector($S$)
      if distance($c, v$) < $\delta$
        $S$ = $S \cup \{v\}$
  until no more word added
  @\textbf{return}@ $S$
\end{lstlisting}
\caption{Keyword Expanding Algorithm}
\label{algo:expand}
\end{figure}

By using vectors to represent keywords and vector distances to measure similarity between keywords, {\tool} can apply $k$-mean, a popular clustering algorithm, to divide a set of keywords into $k$ subsets. Figure~\ref{algo:cluster} illustrates this algorithm. Initially, it randomly selects $k$ \emph{centroids}, each is a vector and represents the center of a cluster. Then, it assigns each keyword $v$ to the cluster whose centroid is closest to $v$ (in term of vector distance). After that, each centroid is re-computed as the mean vector over all keywords assigned to the corresponding cluster. This process repeats until all clusters are stable, i.e. no word is assigned to a cluster different from its current assignment. Because the centroid is the mean vector of each cluster, it could be considered to represent the ``common meaning'' of that cluster. Therefore, by assigning a keyword to the cluster with closest centroid, each keyword is considered to be assigned to a cluster most similar to it.

	

Figure~\ref{algo:expand} illustrates our keyword expanding algorithm. Although it performs the task opposite to keyword clustering, it reuses some ideas of the keyword clustering algorithm. That is, given a set of keywords $S$ for expanding, our algorithm first computes the mean vector $c$ over all keywords in $S$ to represent that set. Then, it adds to $S$ any keyword $v$ that is close enough to $c$, i.e. the distance between $c$ and $v$ is smaller than a pre-defined threshold $\delta$. After that, the mean vector $c$ is re-computed and the process repeats until no new keywords are added. $S$ is then returned with newly added keywords.

\section{Review Search and Trend Analysis}
\label{sec:trend}

Once an analyst has selected a suitable (i.e. as concrete and comprehensive as possible) set of keywords to describe her interests, {\tool} can search for the actual user reviews that are most relevant to those keywords. This is the standard information retrieval task and thus, {\tool} employs the popular \emph{tf.idf} weighting scheme and the Vector Space Model (VSM) for this task~\cite{manning2008introductionInfoRetrieval}. In addition to review search, {\tool} can visualize and analyze the occurrences of the selected keywords over time, helping the analyst to spot any unusual patterns or abnormalities in those occurrences. Let us discuss those two tasks in details.

\subsection{Review Search}

The main goal of the Review Search function is to provide the analyst the actual reviews that are the most relevant to his interests specified by the keywords she provides. To do that, we need a model to represent the reviews and measure the relevance between them and the provided keywords. Vector Space Model~\cite{manning2008introductionInfoRetrieval} is wide-used to represent textual documents in information retrieval systems. In Vector Space Model, a document (e.g. a review) or a query (i.e. a set of keywords) is represented by a high dimensional vector of which each element corresponds to a term or keyword in the vocabulary. Then, the relevance between a document and a query is often computed as the cosine between their vectors.

{\tool} employs Vector Space Model for its Review Search function. It represents each review by a vector and uses the \emph{tf.idf} (term frequency - inverse document frequency~\cite{manning2008introductionInfoRetrieval}), a standard term weighting scheme to compute the element values for those vectors. The term frequency (tf) of a keyword in a review is the number of its occurrences in that review, while the document frequency (df) of a keyword is the number of reviews in the dataset containing it. We compute the tf.idf weight for a keyword as: $$\text{tf.idf} = \frac {\text{tf}} {\log(\text{df}+1)}$$

For example, if we just consider three reviews in Table~\ref{tab:example.query}, keyword \code{``battery''} appears in all of them and appears four times in the last review. Therefore, its term frequency for the last review is 4 and its document frequency (in the whole dataset) is 3. Thus, its tf.idf weight in vector for the last review is 4/(log(3+1)) = 2. Similarly, \code{``drain''} occurs in all three reviews but just once in the last review, thus, its tf.idf weight is 1/(log(3+1)) = 0.5.

{\tool} pre-computes the tf.idf weights and the vectors for all reviews when it extracts keywords from them (see Section~\ref{sec:preprocessing}). When a set of keywords is given for review search, it computes the tf.idf vector for that set (term frequencies are set as 1 and document frequencies are taken from the review data). Then, it computes the cosine of the tf.idf vector of the query with the tf.idf vectors of all available reviews, ranks the reviews by those scores (the higher the more relevant), and presents the results back to the analyst.

\subsection{Trend Analysis}\label{sec:evalts}

	\begin{figure*}[t]
	\centering
	\includegraphics[width=\textwidth]{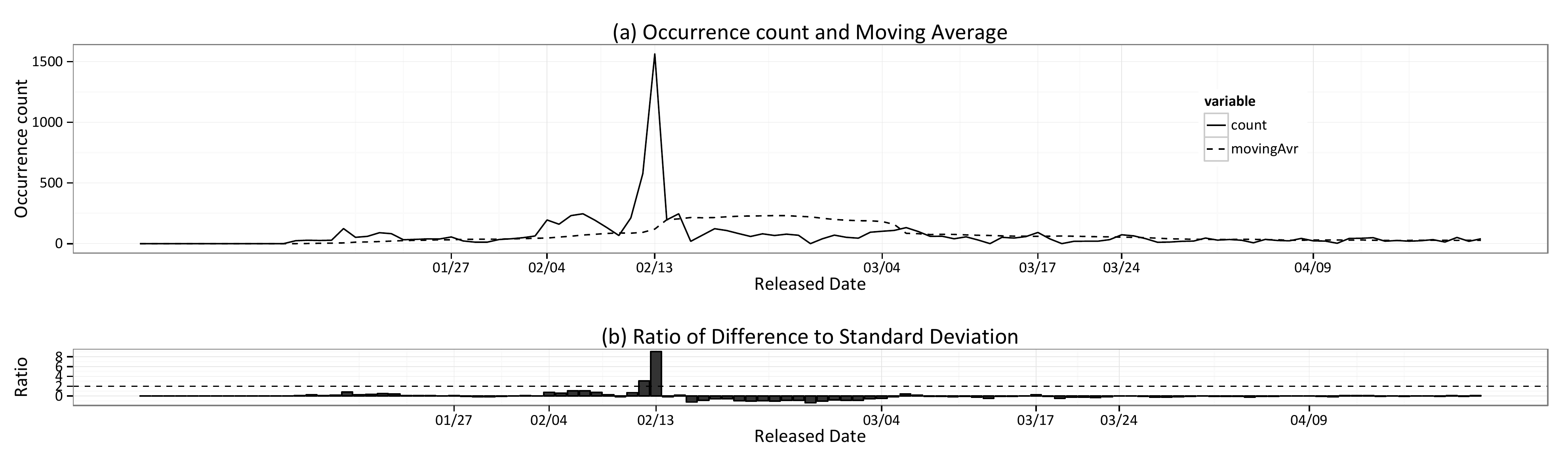}
	\caption{Trend Analysis for ``energy consumption'' issue in Facebook Messenger. The keywords used are: \emph{heat}, \emph{hog}, \emph{usage}, \emph{consumption}, \emph{consume}, \emph{battery}, \emph{drain}, \emph{hogger}, \emph{overheat}, \emph{eater}, \emph{eat}, \emph{drainer}, \emph{power}. Threshold for ratio is 2.}
	\label{fig:trendeval}
	\end{figure*}

Prior studies~\cite{ARminer, Wiscom} suggest that unusual patterns in user reviews often indicate severe, wide-spread issues or problems that make many users unsatisfied. Trend Analysis function in {\tool} allows review analysts to track user reviews on a topic over time and detect any abnormalities in their discussions. This can help app developers to spot issues related their apps early and address the occurred problems in a timely manner.

Trend Analysis is designed based on time series analysis techniques. Once an analyst describes her interest via a set of keywords and a specified time range, {\tool} computes the total occurrences of those keywords for any day in that range to produce a time series. Figure~\ref{fig:trendeval}(a) shows such a time series for the set of keywords related to \emph{``energy consumption''} specified in Table~\ref{tab:example.expand} in the time range from Jan 1, 2015 to Apr 24, 2015. We could see an unusual pike in the occurrences of those keywords around Feb 13, 2015. Minor fluctuations also appear in other days, e.g. around Feb 4, 2015 or Mar 4, 2015.

Such minor fluctuations could be considered as noise. To smooth out such noises and focus on major changes, {\tool} employs simple moving average (SMA), a simple but effective time series analysis technique. Given the time series $x_i, i = 1..n$ ($x_i$ is the total occurrences of the keywords of interest in day $i$) and a lag parameter $k$, the $k$-day simple moving average of time series $x_i$ is another time series $a_i$ where $$a_i = \sum_{j=i-k}^{i-1}x_j$$ In other words, the simple moving average of day $i$ is the average of the last $k$ days before day $i$.

Figure~\ref{fig:trendeval}(a) also shows the 113-days simple moving average of the aforementioned time series for the \emph{``energy consumption''} keywords. We could see that the simple moving average (the dashed line) is much smoother than the original time series (the solid line). That is because moving averaging acts like a low-pass filter. When averaging the original values via a sliding window, the short-term fluctuations in the original values cancel each other.

We could also see that sudden changes in the original time series could not be detected in the simple moving average series alone (they are actually smoothed out too). However, when comparing the original series and its simple moving average, we could see that a major change occurs when the original value is significant higher than the corresponding simple moving average value. Thus, to detect such changes, we define the \emph{relative deviation score} as: $$\tau_i = \frac {x_i - a_i} {\sigma_i}$$ In this formula, $x_i$, $a_i$, and $\sigma_i$ are the corresponding the actual occurrence count, the moving average value, and the standard deviation of the moving average series at day $i$. $a_i$ could be considered as the expected count for day $i$. So $\tau_i$ indicates how far the actual count differs from the expected count relative to the typical difference (i.e. the standard deviation). In statistics, a relative deviation of 2 (i.e. an actual difference of two standard deviations) is often considered as significant. Therefore, if $\tau_i > 2$, {\tool} will report day $i$ to have a sudden, significant change in the trend for the corresponding keywords. Of course, it also allows users to use a different threshold.

Figure~\ref{fig:trendeval}(b) plots the relative deviation score over time for the \emph{``energy consumption''} keywords. As seen in the figure, it is lower than 2 most of the time, except two days Feb 12 and 13, 2015. Those two days are of course the time of a burst in occurrences of those keywords. 
\section{Empirical Evaluation}
\label{sec:eval}
In this section, we present and discuss our empirical evaluation of {\tool} on a large dataset of more than two millions reviews collected from Google Play. Our empirical evaluation focuses on the runtime parameters and the final accuracy of {\tool} in its keyword extraction and recommendation tasks. We also evaluate the correctness and potential usefulness of review search tasks.

\subsection{Data Collection}
\label{sec:data}
\begin{table}
  \centering\sf
  \caption{Statistics of collected data}
    \begin{adjustbox}{max width=0.5\paperwidth}
    \begin{tabular}{lr}
    \toprule
    Total number of mobile apps & 95 \\
    Total number of crawled reviews & 2,106,605 \\
    Average number of reviews per app & 22,174 \\
    Max number of reviews (Clash Of Clans) & 302,936  \\
    Min number of reviews (Amazon MP3) & 1,001  \\
   	Average words per review & 11.1  \\
    Collection Time & Jan 1, 2015 to May 1, 2015\\
    \bottomrule
    \end{tabular}%
    \end{adjustbox}
  \label{tab:data}%
\end{table}%

Table \ref{tab:data} summarizes the dataset we prepared for our evaluation. In total, we have crawled two millions of reviews from 95 mobile apps on Google Play from January 1, 2015 to May 1, 2015. Each crawled review contains a title, a long text description of the review content, the creation time, the reviewer ID, and the associated rating.

The apps chosen for our study are among the most popular apps in this market such as Facebook, Twitter, WhatsApp, Snapchat, Viber, Instagram, and Clash of Clans. The full list of apps can be accessed at {\tool}'s website~\url{http://useal.cs.usu.edu/mark}. On average, each app has about 20 thousands reviews. However, the actual number of crawled reviews varies between apps because some of them have more active users than others. For example, Clash of Clans is a very popular game with near 30 millions of daily active users~\cite{cocstatistic}. Within five months of our crawling process, we have obtained more than 300 thousands reviews for that app.

Because we use an open-source crawler~\cite{gplayCrawler} and Google Play returns only 500 reviews for each request, we had to crawl the reviews of the chosen apps continuously overtime to obtain to the most recent ones. However, the crawled reviews might not be all available reviews because the Google Play APIs we use for crawling do not have any option for storing the current states or excluding the previously crawled results.



\subsection{Keyword Extraction}
\subsubsection{Non-English reviews classifier}
\begin{table*}[t]
  \centering
  \sf
  \caption{Examples of Uni-gram and Bi-gram ratios on Detecting Non-English reviews}
    \begin{adjustbox}{max width=0.85\paperwidth}
    \begin{tabular}{lllll}
    \toprule
    \textbf{Word Count} & \textbf{Bi-gram Ratio} & \textbf{Uni-gram Ratio} & \textbf{ Reviews} & \textbf{Note} \bigstrut\\
    \midrule
    12    & 0.58 & 0.67 & \parbox[t]{.6\linewidth}{it's so fun i'm gong to die.utut77brhrhuvrivriy7vht7hrtour7} & Minor mistakes \bigstrut[t]\\
    59    & 0.58 & 0.68 & \parbox[t]{.6\linewidth}{es, incomparable.this perhas one, if not, the best radio stations i've tuned into. has variety and though i haven't yet figures it all out how to scroll through the stations, record and everything, i like it. si tubiera que recomendar musica de internet, definitivamente esta seria mi primera opcion. en tu idioma, canciones retro y variedad chingado. es inigualabe} & Written in two languages \\
    16    & 0.56 & 0.69 & \parbox[t]{.6\linewidth}{need support group! clans anonymous!.this game is completely addictive! i....neeed.....my......nexxxttt......doooossseeee l....offfff...thiiiiiisssss....} & Misspelled on purpose \\
    15    & 0.38   & 0.67 & \parbox[t]{.6\linewidth}{.nce n i lov dis app.... n hlps tu clear de threats wch wre thre in ma phne.} & Too many slangs \\

    \midrule
    20    & 0.4   & 0.55  & \parbox[t]{.6\linewidth}{.maganda kasi disente pwede sa mga kids walang bastos d tulad ng ibang larong on-line..long live clash of clans!!!!!!!} & Foreign language\bigstrut[t]\\
    5     & 0.4   & 0.6   & \parbox[t]{.6\linewidth}{buddha hong lien hoa.ommanipadmehum?!:-)...."} & Foreign language\\
    \bottomrule
    \end{tabular}%
      \end{adjustbox}
  \label{tab:nonengfilter}%
\end{table*}%
We use the dataset of 400 reviews we have manually labeled in a preliminary analysis (see Section II) to evaluate the accuracy of our English review classifier. This dataset contains 245 English reviews and 155 non-English ones. In our experiment, we varied the thresholds of English uni-gram and bi-gram ratios from 0 to 1 with the increment of 0.01 and used those thresholds to classify and compute the classification accuracy. Finally, the best accuracy of 86.5\% was obtained with the uni-gram and bi-gram thresholds of 0.64 and 0.38, respectively. We consider this level of accuracy reasonable. However, in future work, we plan to investigate in more sophisticated classification techniques to achieve higher levels of accuracy.

Table~\ref{tab:nonengfilter} lists some examples of English (in the upper half) and non-English (in the lower half) reviews classified by our technique with the corresponding uni-gram and bi-gram ratios for some of our examples. As seen in the table, our technique can recognize English reviews with words written in another language or even unrecognizable.

\subsubsection{Customized Stemming}\label{sec:stemeval}

To evaluate our customized stemmer, we prepared a dataset by randomly selecting 1,000 words from our review data. This dataset includes both verbs and nouns which are tagged by the Stanford PoS Tagger. We manually stemmed those words to their respective base forms. Then we ran our stemmer on this dataset and computed the accuracy as the percentage of words whose base forms were derived by our stemmer match the base forms we have manually stemmed. We also run the widely used Stanford Lemmatizer tool for comparison purpose. 

Overall, our stemmer was able to stem correctly 97.9\% of the words while the Stanford Lemmatizer can only got 90.8\% right. The main contributor for that improvement is likely the domain specific dictionary of misspelled words and special names (see in Section \ref{sec:misspell}) included in our stemming and spelling correction procedure.

\subsection{Keywords Recommendation}
\subsubsection{Keywords Clustering}\label{sec:grouping}

\begin{table*}[t]

  \centering\sf

  	  \renewcommand{\arraystretch}{0.6}
  \caption{K-mean clustering result for top 100 highest ranked keywords. Each word is accompanied by its acceptance rate of 8 researchers (x/8). Accuracy is the average of the acceptance rates.}
  \begin{adjustbox}{max width=\textwidth}
    \setlength{\tabcolsep}{2pt}
       \begin{tabular}{|r|r|r|r|r|r|r|r|r|r|r|r|r|r|r|r|r|r|r|}
           \toprule
           \textbf{Main concerns} & \multicolumn{2}{c|}{\textbf{Battery \& versioning}} & \multicolumn{2}{c|}{\textbf{Connection}} & \multicolumn{2}{c|}{\textbf{Unrecoverable error}} & \multicolumn{2}{c|}{\textbf{Messaging}} & \multicolumn{4}{c|}{\textbf{Snapchat}} & \multicolumn{4}{c|}{\textbf{Authentication}} & \multicolumn{2}{c|}{\textbf{Facebook}} \bigstrut\\
           \midrule
           \multicolumn{1}{|c|}{\multirow{13}[2]{*}{}} & \multicolumn{1}{r}{drain} & \multicolumn{1}{l|}{(8/8)} & \multicolumn{1}{r}{data} & \multicolumn{1}{l|}{(7/8)} & \multicolumn{1}{r}{crash} & \multicolumn{1}{l|}{(8/8)} & \multicolumn{1}{r}{message} & \multicolumn{1}{l|}{(8/8)} & \multicolumn{1}{r}{update} & \multicolumn{1}{l}{(8/8)} & \multicolumn{1}{r}{dislike} & \multicolumn{1}{l|}{(5/8)} & \multicolumn{1}{r}{fix} & \multicolumn{1}{l}{(8/8)} & \multicolumn{1}{r}{login} & \multicolumn{1}{l|}{(8/8)} & \multicolumn{1}{r}{feed} & \multicolumn{1}{l|}{(8/8)} \bigstrut[t]\\
           \multicolumn{1}{|c|}{} & \multicolumn{1}{r}{ruin} & \multicolumn{1}{l|}{(8/8)} & \multicolumn{1}{r}{connection} & \multicolumn{1}{l|}{(8/8)} & \multicolumn{1}{r}{open} & \multicolumn{1}{l|}{(6/8)} & \multicolumn{1}{r}{send} & \multicolumn{1}{l|}{(8/8)} & \multicolumn{1}{r}{story} & \multicolumn{1}{l}{(7/8)} & \multicolumn{1}{r}{} & \multicolumn{1}{l|}{} & \multicolumn{1}{r}{log} & \multicolumn{1}{l}{(7/8)} & \multicolumn{1}{r}{fail} & \multicolumn{1}{l|}{(7/8)} & \multicolumn{1}{r}{refresh} & \multicolumn{1}{l|}{(8/8)} \\
           \multicolumn{1}{|c|}{} & \multicolumn{1}{r}{bug} & \multicolumn{1}{l|}{(8/8)} & \multicolumn{1}{r}{retry} & \multicolumn{1}{l|}{(8/8)} & \multicolumn{1}{r}{freeze} & \multicolumn{1}{l|}{(8/8)} & \multicolumn{1}{r}{notification} & \multicolumn{1}{l|}{(8/8)} & \multicolumn{1}{r}{load} & \multicolumn{1}{l}{(8/8)} & \multicolumn{1}{r}{} & \multicolumn{1}{l|}{} & \multicolumn{1}{r}{uninstall} & \multicolumn{1}{l}{(8/8)} & \multicolumn{1}{r}{delete} & \multicolumn{1}{l|}{(7/8)} & \multicolumn{1}{r}{newsfeed} & \multicolumn{1}{l|}{(8/8)} \\
           \multicolumn{1}{|c|}{} & \multicolumn{1}{r}{break} & \multicolumn{1}{l|}{(6/8)} & \multicolumn{1}{r}{permission} & \multicolumn{1}{l|}{(6/8)} & \multicolumn{1}{r}{close} & \multicolumn{1}{l|}{(8/8)} & \multicolumn{1}{r}{code} & \multicolumn{1}{l|}{(6/8)} & \multicolumn{1}{r}{video} & \multicolumn{1}{l}{(8/8)} & \multicolumn{1}{r}{} & \multicolumn{1}{l|}{} & \multicolumn{1}{r}{expire} & \multicolumn{1}{l}{(8/8)} & \multicolumn{1}{r}{respond} & \multicolumn{1}{l|}{(2/8)} & \multicolumn{1}{r}{news} & \multicolumn{1}{l|}{(8/8)} \\
           \multicolumn{1}{|c|}{} & \multicolumn{1}{r}{version} & \multicolumn{1}{l|}{(5/8)} & \multicolumn{1}{r}{turn} & \multicolumn{1}{l|}{(3/8)} & \multicolumn{1}{r}{stop} & \multicolumn{1}{l|}{(8/8)} & \multicolumn{1}{r}{band} & \multicolumn{1}{l|}{(1/8)} & \multicolumn{1}{r}{screen} & \multicolumn{1}{l}{(6/8)} & \multicolumn{1}{r}{} & \multicolumn{1}{l|}{} & \multicolumn{1}{r}{session} & \multicolumn{1}{l}{(8/8)} & \multicolumn{1}{r}{say} & \multicolumn{1}{l|}{(6/8)} & \multicolumn{1}{r}{scroll} & \multicolumn{1}{l|}{(7/8)} \\
           \multicolumn{1}{|c|}{} & \multicolumn{1}{r}{disgust} & \multicolumn{1}{l|}{(6/8)} & \multicolumn{1}{r}{network} & \multicolumn{1}{l|}{(8/8)} & \multicolumn{1}{r}{restart} & \multicolumn{1}{l|}{(8/8)} & \multicolumn{1}{r}{bar} & \multicolumn{1}{l|}{(2/8)} & \multicolumn{1}{r}{snap} & \multicolumn{1}{l}{(8/8)} & \multicolumn{1}{r}{} & \multicolumn{1}{l|}{} & \multicolumn{1}{r}{error} & \multicolumn{1}{l}{(8/8)} & \multicolumn{1}{r}{download} & \multicolumn{1}{l|}{(7/8)} & \multicolumn{1}{r}{post} & \multicolumn{1}{l|}{(8/8)} \\
           \multicolumn{1}{|c|}{} & \multicolumn{1}{r}{lag} & \multicolumn{1}{l|}{(8/8)} & \multicolumn{1}{r}{wifi} & \multicolumn{1}{l|}{(8/8)} & \multicolumn{1}{r}{shut} & \multicolumn{1}{l|}{(8/8)} & \multicolumn{1}{r}{receive} & \multicolumn{1}{l|}{(8/8)} & \multicolumn{1}{r}{upload} & \multicolumn{1}{l}{(8/8)} & \multicolumn{1}{r}{} & \multicolumn{1}{l|}{} & \multicolumn{1}{r}{reinstall} & \multicolumn{1}{l}{(8/8)} & \multicolumn{1}{r}{password} & \multicolumn{1}{l|}{(7/8)} & \multicolumn{1}{r}{page} & \multicolumn{1}{l|}{(7/8)} \\
           \multicolumn{1}{|c|}{} & \multicolumn{1}{r}{lollipop} & \multicolumn{1}{l|}{(6/8)} & \multicolumn{1}{r}{server} & \multicolumn{1}{l|}{(8/8)} & \multicolumn{1}{r}{annoy} & \multicolumn{1}{l|}{(7/8)} & \multicolumn{1}{r}{email} & \multicolumn{1}{l|}{(8/8)} & \multicolumn{1}{r}{view} & \multicolumn{1}{l}{(8/8)} & \multicolumn{1}{r}{} & \multicolumn{1}{l|}{} & \multicolumn{1}{r}{install} & \multicolumn{1}{l}{(7/8)} & \multicolumn{1}{r}{account} & \multicolumn{1}{l|}{(8/8)} & \multicolumn{1}{r}{read} & \multicolumn{1}{l|}{(7/8)} \\
           \multicolumn{1}{|c|}{} & \multicolumn{1}{r}{downgrade} & \multicolumn{1}{l|}{(8/8)} & \multicolumn{1}{r}{} & \multicolumn{1}{l|}{} & \multicolumn{1}{r}{disappear} & \multicolumn{1}{l|}{(8/8)} & \multicolumn{1}{r}{access} & \multicolumn{1}{l|}{(7/8)} & \multicolumn{1}{r}{snapchat} & \multicolumn{1}{l}{(8/8)} & \multicolumn{1}{r}{} & \multicolumn{1}{l|}{} & \multicolumn{1}{r}{sign} & \multicolumn{1}{l}{(7/8)} & \multicolumn{1}{r}{refuse} & \multicolumn{1}{l|}{(8/8)} & \multicolumn{1}{r}{show} & \multicolumn{1}{l|}{(8/8)} \\
           \multicolumn{1}{|c|}{} & \multicolumn{1}{r}{tab} & \multicolumn{1}{l|}{(3/8)} & \multicolumn{1}{r}{} & \multicolumn{1}{l|}{} & \multicolumn{1}{r}{pop} & \multicolumn{1}{l|}{(5/8)} & \multicolumn{1}{r}{verification} & \multicolumn{1}{l|}{(7/8)} & \multicolumn{1}{r}{mess} & \multicolumn{1}{l}{(4/8)} & \multicolumn{1}{r}{} & \multicolumn{1}{l|}{} & \multicolumn{1}{r}{happen} & \multicolumn{1}{l}{(3/8)} & \multicolumn{1}{r}{reset} & \multicolumn{1}{l|}{(7/8)} & \multicolumn{1}{r}{click} & \multicolumn{1}{l|}{(2/8)} \\
           \multicolumn{1}{|c|}{} & \multicolumn{1}{r}{pos} & \multicolumn{1}{l|}{(3/8)} & \multicolumn{1}{r}{} & \multicolumn{1}{l|}{} & \multicolumn{1}{r}{minute} & \multicolumn{1}{l|}{(2/8)} & \multicolumn{1}{r}{verify} & \multicolumn{1}{l|}{(7/8)} & \multicolumn{1}{r}{skip} & \multicolumn{1}{l}{(5/8)} & \multicolumn{1}{r}{} & \multicolumn{1}{l|}{} & \multicolumn{1}{r}{waste} & \multicolumn{1}{l}{(6/8)} & \multicolumn{1}{r}{} & \multicolumn{1}{l|}{} & \multicolumn{1}{r}{timeline} & \multicolumn{1}{l|}{(8/8)} \\
           \multicolumn{1}{|c|}{} & \multicolumn{1}{r}{space} & \multicolumn{1}{l|}{(4/8)} & \multicolumn{1}{r}{} & \multicolumn{1}{l|}{} & \multicolumn{1}{r}{reason} & \multicolumn{1}{l|}{(1/8)} & \multicolumn{1}{r}{} & \multicolumn{1}{l|}{} & \multicolumn{1}{r}{bestfriend} & \multicolumn{1}{l}{(6/8)} & \multicolumn{1}{r}{} & \multicolumn{1}{l|}{} & \multicolumn{1}{r}{try} & \multicolumn{1}{l}{(7/8)} & \multicolumn{1}{r}{} & \multicolumn{1}{l|}{} & \multicolumn{1}{r}{comment} & \multicolumn{1}{l|}{(8/8)} \\
           \multicolumn{1}{|c|}{} & \multicolumn{1}{r}{} & \multicolumn{1}{l|}{} & \multicolumn{1}{r}{} & \multicolumn{1}{l|}{} & \multicolumn{1}{r}{hang} & \multicolumn{1}{l|}{(8/8)} & \multicolumn{1}{r}{} & \multicolumn{1}{l|}{} & \multicolumn{1}{r}{discover} & \multicolumn{1}{l}{(5/8)} & \multicolumn{1}{r}{} & \multicolumn{1}{l|}{} & \multicolumn{1}{r}{attempt} & \multicolumn{1}{l}{(7/8)} & \multicolumn{1}{r}{} & \multicolumn{1}{l|}{} & \multicolumn{1}{r}{middle} & \multicolumn{1}{l|}{(1/8)} \bigstrut[b]\\
           \midrule
           \textbf{Accuracy} & \multicolumn{2}{c|}{\textbf{76.04\%}} & \multicolumn{2}{c|}{\textbf{87.50\%}} & \multicolumn{2}{c|}{\textbf{81.73\%}} & \multicolumn{2}{c|}{\textbf{79.55\%}} & \multicolumn{4}{c|}{\textbf{83.93\%}} & \multicolumn{4}{c|}{\textbf{86.41\%}} & \multicolumn{2}{c|}{\textbf{84.62\%}} \bigstrut\\
           \midrule
           \textbf{Avr. Accuracy} & \multicolumn{18}{c|}{\textbf{83.11\%}} \bigstrut\\
           \bottomrule
           \end{tabular}%

      \end{adjustbox}
  \label{tab:clustereval}%
\end{table*}%

To evaluate our Keyword Clustering technique, we selected the top 100 keywords ranked based on our contrast score. Then, we used the simple formula $k=\sqrt{n/2}$ to estimate the number of clusters (seven in this case). Table~\ref{tab:clustereval} shows the results when {\tool} clustered our selected keywords into seven clusters. 
 
We asked a team of eight Computer Science researchers to manually label those clusters and identify irrelevant words in each cluster. Then, we discussed to choose the clusters' labels and combined the results. In Table~\ref{tab:clustereval}, the acceptance rate of each word is given next to it. For example, all eight researchers considered \code{``crash''} relevant for the issue of \emph{``unrecoverable error''}, while only two considered \code{``minute''} relevant.

We measure the accuracy of each cluster as the average acceptance rate of its words. For example, cluster \emph{``connection''} has its words well accepted, thus, has an accuracy of 87.5\%. However, cluster \emph{``battery \& versioning''} is worse with an accuracy of 76\%. The overall accuracy, as the average over all clusters, is 83\%.

It is interesting that Facebook and Snapchat have their own clusters from top 100 most negative words. Our manual investigation of their recent releases suggests that they often have problems with such updates. Due to their popularity, i.e. having much higher number of active users than other apps, such problems lead to high amounts of negative reviews which could partially explain why many of their keywords appeared on the top negative list. This result suggests that their developers should be more careful in quality control of their future releases.

\subsubsection{Keywords Expanding}

\begin{table*}[t]
  \centering\sf
  \caption{Keyword expanding results for several popular topics from Wiscom\cite{Wiscom}. Each word is accompanied by its acceptance rate of 8 researchers (x/8). Accuracy is the average of the acceptance rates. Chosen threshold is 0.75}
  \begin{adjustbox}{max width=\textwidth}
  \setlength{\tabcolsep}{2pt}
  	  \renewcommand{\arraystretch}{0.6}
    \begin{tabular}{|r|r|r|r|r|r|r|r|r|r|r|r|r|r|r|r|r|}
    \toprule
    \textbf{Starter word} & \multicolumn{4}{c|}{\textbf{crash}} & \multicolumn{2}{c|}{\textbf{compatibility}} & \multicolumn{2}{c|}{\textbf{connection}} & \multicolumn{2}{c|}{\textbf{pay}} & \multicolumn{2}{c|}{\textbf{call}} & \multicolumn{2}{c|}{\textbf{camera}} & \multicolumn{2}{c|}{\textbf{ads}} \bigstrut\\
    \midrule
    \multicolumn{1}{|c|}{\multirow{10}[2]{*}{}} & \multicolumn{1}{r}{reboot} & \multicolumn{1}{l}{(8/8)} & \multicolumn{1}{r}{close} & \multicolumn{1}{l|}{(8/8)} & \multicolumn{1}{r}{android} & \multicolumn{1}{l|}{(7/8)} & \multicolumn{1}{r}{wifi} & \multicolumn{1}{l|}{(8/8)} & \multicolumn{1}{r}{buy} & \multicolumn{1}{l|}{(8/8)} & \multicolumn{1}{r}{voice} & \multicolumn{1}{l|}{(8/8)} & \multicolumn{1}{r}{record} & \multicolumn{1}{l|}{(7/8)} & \multicolumn{1}{r}{commercia} & \multicolumn{1}{l|}{(6/8)} \bigstrut[t]\\
    \multicolumn{1}{|c|}{} & \multicolumn{1}{r}{shut} & \multicolumn{1}{l}{(8/8)} & \multicolumn{1}{r}{open} & \multicolumn{1}{l|}{(6/8)} & \multicolumn{1}{r}{} & \multicolumn{1}{l|}{} & \multicolumn{1}{r}{network} & \multicolumn{1}{l|}{(8/8)} & \multicolumn{1}{r}{spend} & \multicolumn{1}{l|}{(8/8)} & \multicolumn{1}{r}{whatsapp} & \multicolumn{1}{l|}{(8/8)} & \multicolumn{1}{r}{zoom} & \multicolumn{1}{l|}{(8/8)} & \multicolumn{1}{r}{advertisement} & \multicolumn{1}{l|}{(8/8)} \\
    \multicolumn{1}{|c|}{} & \multicolumn{1}{r}{hang} & \multicolumn{1}{l}{(8/8)} & \multicolumn{1}{r}{} & \multicolumn{1}{l|}{} & \multicolumn{1}{r}{} & \multicolumn{1}{l|}{} & \multicolumn{1}{r}{4g} & \multicolumn{1}{l|}{(8/8)} & \multicolumn{1}{r}{purchase} & \multicolumn{1}{l|}{(7/8)} & \multicolumn{1}{r}{viber} & \multicolumn{1}{l|}{(8/8)} & \multicolumn{1}{r}{cam} & \multicolumn{1}{l|}{(8/8)} & \multicolumn{1}{r}{advert} & \multicolumn{1}{l|}{(8/8)} \\
    \multicolumn{1}{|c|}{} & \multicolumn{1}{r}{restart} & \multicolumn{1}{l}{(8/8)} & \multicolumn{1}{r}{} & \multicolumn{1}{l|}{} & \multicolumn{1}{r}{} &       & \multicolumn{1}{r}{3g} & \multicolumn{1}{l|}{(8/8)} & \multicolumn{1}{r}{money} & \multicolumn{1}{l|}{(8/8)} & \multicolumn{1}{r}{cal} & \multicolumn{1}{l|}{(4/8)} & \multicolumn{1}{r}{flash} & \multicolumn{1}{l|}{(8/8)} & \multicolumn{1}{r}{spoor} & \multicolumn{1}{l|}{(0/8)} \\
    \multicolumn{1}{|c|}{} & \multicolumn{1}{r}{exit} & \multicolumn{1}{l}{(7/8)} & \multicolumn{1}{r}{} & \multicolumn{1}{l|}{} & \multicolumn{1}{r}{} & \multicolumn{1}{l|}{} & \multicolumn{1}{r}{connectivity} & \multicolumn{1}{l|}{(8/8)} & \multicolumn{1}{r}{earn} & \multicolumn{1}{l|}{(6/8)} & \multicolumn{1}{r}{skype} & \multicolumn{1}{l|}{(8/8)} & \multicolumn{1}{r}{} & \multicolumn{1}{l|}{} & \multicolumn{1}{r}{advertise} & \multicolumn{1}{l|}{(8/8)} \\
    \multicolumn{1}{|c|}{} & \multicolumn{1}{r}{freeze} & \multicolumn{1}{l}{(8/8)} & \multicolumn{1}{r}{} & \multicolumn{1}{l|}{} & \multicolumn{1}{r}{} & \multicolumn{1}{l|}{} & \multicolumn{1}{r}{2g} & \multicolumn{1}{l|}{(8/8)} & \multicolumn{1}{r}{} & \multicolumn{1}{l|}{} & \multicolumn{1}{r}{tango} & \multicolumn{1}{l|}{(7/8)} & \multicolumn{1}{r}{} &       & \multicolumn{1}{r}{} &  \\
    \multicolumn{1}{|c|}{} & \multicolumn{1}{r}{stop} & \multicolumn{1}{l}{(8/8)} & \multicolumn{1}{r}{} & \multicolumn{1}{l|}{} & \multicolumn{1}{r}{} & \multicolumn{1}{l|}{} & \multicolumn{1}{r}{lte} & \multicolumn{1}{l|}{(8/8)} & \multicolumn{1}{r}{} & \multicolumn{1}{l|}{} & \multicolumn{1}{r}{facility} & \multicolumn{1}{l|}{(3/8)} & \multicolumn{1}{r}{} & \multicolumn{1}{l|}{} & \multicolumn{1}{r}{} &  \\
    \multicolumn{1}{|c|}{} & \multicolumn{1}{r}{load} & \multicolumn{1}{l}{(5/8)} & \multicolumn{1}{r}{} & \multicolumn{1}{l|}{} & \multicolumn{1}{r}{} & \multicolumn{1}{l|}{} & \multicolumn{1}{r}{signal} & \multicolumn{1}{l|}{(8/8)} & \multicolumn{1}{r}{} &       & \multicolumn{1}{r}{} &       & \multicolumn{1}{r}{} &       & \multicolumn{1}{r}{} &  \\
    \multicolumn{1}{|c|}{} & \multicolumn{1}{r}{reopen} & \multicolumn{1}{l}{(8/8)} & \multicolumn{1}{r}{} & \multicolumn{1}{l|}{} & \multicolumn{1}{r}{} & \multicolumn{1}{l|}{} & \multicolumn{1}{r}{internet} & \multicolumn{1}{l|}{(8/8)} & \multicolumn{1}{r}{} & \multicolumn{1}{l|}{} & \multicolumn{1}{r}{} &       & \multicolumn{1}{r}{} & \multicolumn{1}{l|}{} & \multicolumn{1}{r}{} &  \\
    \multicolumn{1}{|c|}{} & \multicolumn{1}{r}{respond} & \multicolumn{1}{l}{(8/8)} & \multicolumn{1}{r}{} & \multicolumn{1}{l|}{} & \multicolumn{1}{r}{} & \multicolumn{1}{l|}{} & \multicolumn{1}{r}{connect} & \multicolumn{1}{l|}{(8/8)} & \multicolumn{1}{r}{} & \multicolumn{1}{l|}{} & \multicolumn{1}{r}{} & \multicolumn{1}{l|}{} & \multicolumn{1}{r}{} & \multicolumn{1}{l|}{} & \multicolumn{1}{r}{} & \multicolumn{1}{l|}{} \bigstrut[b]\\
    \midrule
    \textbf{Accuracy} & \multicolumn{4}{c|}{\textbf{93.75\%}} & \multicolumn{2}{c|}{\textbf{87.5\%}} & \multicolumn{2}{c|}{\textbf{93.18\%}} & \multicolumn{2}{c|}{\textbf{92.5\%}} & \multicolumn{2}{c|}{\textbf{82.14\%}} & \multicolumn{2}{c|}{\textbf{96.88\%}} & \multicolumn{2}{c|}{\textbf{75\%}} \bigstrut\\
    \midrule
    \textbf{Avr. Accuracy} & \multicolumn{16}{c|}{\textbf{89.72\%}} \bigstrut\\
    \bottomrule
    \end{tabular}%

  \end{adjustbox}
  \label{tab:discoverwords}%
\end{table*}%

In this experiment, we asked {\tool} to expand keywords from review topics reported previously~\cite{Wiscom}. To simplify the experiment, we selected only one keyword of each topic for the initial set. Table~\ref{tab:discoverwords} shows the expanding results. Similar to the previous experiment, we also asked the team of eight researchers to evaluate these results in the same manner, i.e. identifying the unrelated keywords. 

The evaluation results in Table~\ref{tab:discoverwords} indicate the overall accuracy of our technique is of 89.7\%. Two topics \emph{``ads''} or \emph{``call''} accidently have low accuracy because of misspelled words like \code{``cal''} or \code{``spoor''}. Nevertheless, the overall result suggests that our expanding technique can capture semantics of a keyword set and add relevant words to it.
 
We ran this experiment with the distance threshold $\delta$ of 0.25 and also ran with thresholds of $0.2$ and lower. However, due to such low thresholds, some topics could not be expanded because {\tool} cannot find words whose similarity measurements could pass those thresholds.

\subsection{Review Search}

\begin{figure}[t]
\centering
\includegraphics[width=0.5\textwidth]{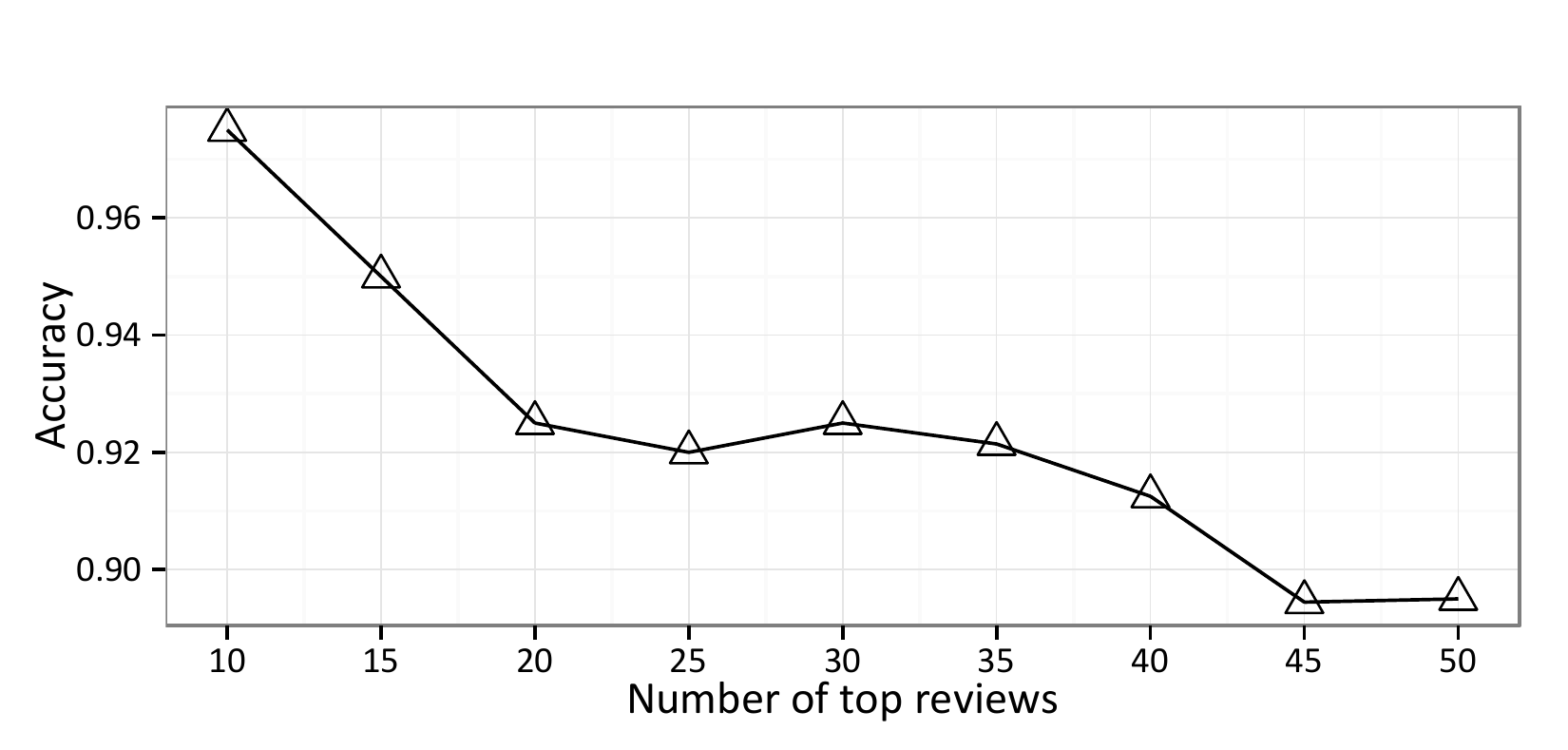}
\caption{Searching accuracy for top returned reviews for battery consumption concern. The keywords for this search are: \emph{heat}, \emph{hog}, \emph{usage}, \emph{consumption}, \emph{consume}, \emph{battery}, \emph{drain}, \emph{hogger}, \emph{overheat}, \emph{eater}, \emph{eat}, \emph{drainer}, \emph{power} }
\label{fig:searchrev}
\end{figure}

We used the same set of keywords about \emph{``energy consumption''} from Section \ref{sec:evalts} to evaluate our review search technique on Facebook Messenger. Our team of four authors read the top 50 returned reviews and manually validated if such a review is relevant to this topic or not.

Figure~\ref{fig:searchrev} shows the accuracy computed for top 10, 15, 20, 25, 30, 45, 50 reviews. As seen in the figure, the overall accuracy is high in the range of 90-97\%. The top 10 reviews are mostly relevant, and the accuracy is no surprise lower with more reviews. Nevertheless, these results suggest that our keyword-based review searching technique can reliably provides reviews relevant to the interests of app developers.

\subsection{Threat of Validity}

The top threat of validity of our study is the fact that our evaluation is performed on datasets manually labeled by the authors and is validated by human subjects. In addition, our subjects are students and professors, while the intended users of our tool should be professional review analysts and app developers. Finally, our datasets and experiments are small in scale (e.g. 95 downloaded apps vs millions available apps) and the apps chosen in our study might not be representative.

To address those threats of validity, we have publicly published our tool, provided our experiment data, and made our work open-source. We encourage fellow researchers to replicate and extend our experiments, send us feedbacks or feature requests, and contribute to our open-sourced tool.

\section{Related Work}\label{sec:relatedwork}
There are a number of empirical and exploratory studies on the importance of app's reviews in app development process.

In \cite{Vasa:preliminary}, Vasa et al. made an exploratory study about how users input their reviews on app stores and what could affect the way they write reviews. Later, Hoon et al.\cite{hoon2013analysisReviewLandscape} analyzed nearly 8 millions reviews on Apple AppStore to discover several statistical characteristics to suggest developers constantly watching for the changes in user's expectations to adapt their apps. Again on Apple App Store, an emprical study about user's feedback was made by Pagano et al.\cite{feedbackEmpirical}. Similarly, Khalid et al. suggest that there are at least 12 types of complaints about iOS apps\cite{Khalid:identifyIOS}. They explored various aspects that influent user reviews such as time of release, topics and several properties including quality and constructiveness to understand their impacts on apps.

Other than reviews, price and rating of apps can also affect how people provide their feedbacks, as suggested in \cite{iacob2013:complain} by Iacob et al. Meanwhile,  Bavota et al.\cite{apiVSRating} studied the relationship between API changes and their faulty level with app ratings. Recently, Martin et al. \cite{martin:appsamplingprob} reported the sampling bias researchers might encounter when mining information from app reviews.

Thus so far, there are very few works in mining useful information from user's reviews on app markets. One of the earliest work is from Chandy et al. \cite{Chandy:Spam} who proposed a classification model for spamming reviews on Apple AppStore using a simple latent model. On the other hand, Carreno et al. \cite{Carreno} extract changes or additional requirements for new versions automatically using information retrieval techniques. Our work is different from theirs in the main goals: We focus on extracting user opinions based on keywords to map the way developers express their concerns to user's way.

More to the mining techniques, in \cite{guzman2014:reviewsAnalysis}, Guzman et al. extracts features from app reviews in form of collocations and summarizes them with Latent Dirichlet Allocation (LDA)\cite{LDA} and their sentiment. In their work, they score the sentiment of reviews by using SentiStrength\cite{SentiStrength}, which is a lexical sentiment approach. The keywords extraction approach in our work is very close to the meaning of features extraction, but with extra flexibility for developers and users to map their expressions. Instead of a collocation of two words, our keywords can come in a set and carries an entirely different dimension of information. We rank our keywords using ratings from users, which is a more domain specific approach for review analysis than Guzman's general lexical sentiment approach.

Another work, which is more closely related to our work, is Wiscom \cite{Wiscom}. Their work on sentiment analysis of words using Linear Regression Model is comparable to our ranking scheme but with a different intention. We try to address the impact of keyword's concerns to users while they want to address the impact on sentiment of a keywords to discover inconsistent review. On "meso level" they use a LDA model to analyze topics of user reviews based on their distribution. Similarly, we group the keywords using K-mean clustering on vector-space representation of words, but our approach focus on exploiting different layers of semantic meaning for words inside the corpus, which give us a different perspective of user opinions. To the best of our knowledge, their work is also the first work to mention the use of timeseries on reviews for analysis, however, their approach was to use root cause analysis based on the observed busts in negative or positive comments, which does not address the problems that may lie inside normal stream of comments. Our approach using keywords can automatically discover changes in trends of given concerns regardless of the total reviews' number.

Interestingly, Iacob et al \cite{Iacob:mara} designed a prototype (MARA) to retrieve app feature requests from comments using a set of linguistic rules. The rules were derived manually from actual text of reviews and the retrieved features are further analyzed using LDA to find common topics among them. This approach shares one similar aim with us: to extract features. However, their feature level stay at the phrase level while our features can be described by keywords, which may be more intuitive for both developers and users.

One of the most recent work in extracting information from reviews is AR-Miner \cite{ARminer}. Chen et al. propose a computational framework to extract and rank informative reviews at sentence level. They adopt the semi-supervised algorithm Expectation Maximization for Naive Bayes (EMNB)\cite{nigam2000text:EMNB} to classify between informative and non-informative reviews. To rank the reviews, they use a ranking schema based on several meta-data of reviews and try to suggest the most informative ones. Our work is different from their work in both purpose and approach: Our purpose is to extract user opinion, while they want to find and rank most informative reviews, so the benefit for developers is different. Moreover, we focus on keywords level because their distributed representations can discover more detailed semantic meanings of user's reviews.

From an empirical study suggested that there are error-prone permissions reported in user reviews, Gomez et al. \cite{buggyappchecker} developed a static error-proneness checker for app based on permissions. This work uses LDA to identify topics in reviews and reported permissions. As other works, it is different from our main purpose and approach of mining keyword's semantical meaning.

Finally, our framework is the first to provide a reliable way to search for most relevant reviews based on keywords approach, which is yet to be mentioned by any prior work.

\section{Conclusion}\label{sec:conclusion}
In this work, we proposed {\tool} as a semi-automated framework to collect and mine user opinions from app markets using a keyword-based approach. We developed and applied several automated, customized techniques for our main tasks, including: extracting keywords from raw reviews, ranking them, grouping them based on their semantic similarity, searching for most relevant reviews to a set of keywords, visualizing their occurrence over time and reporting any unusual patterns.
	
	From our observations on the difficulties of processing raw reviews, we proposed an original technique to classify non-English ones and developed a customized stemmer to normalize their keywords. Our evaluations show that the classifier is able to correctly label 86.5\% reviews in our test set. Our customized stemmer is also proved to be of higher accuracy for stemming app reviews data than the general purpose lematization tool from Stanford for our test set.
	
	Exploiting the multi-degree semantical meaning property of distributed vector-space representation of words, our clustering and expanding techniques can discover highly related keywords that express common concerns. Our case studies show that these techniques is able to reach 83\% accuracy for grouping and 89.7\% accuracy for expanding tasks.
	
	Using a set of keywords discovered from the above steps, {\tool} can help developers to search for most relevance reviews to that set. In our case study of searching for \emph{battery consumption} concern in Facebook Messenger, we found that 90\% of top 50 returned reviews satisfy the query.
	
	In the tasks of discovering abnormalities of keywords occurrence in a time period, our analysis technique utilizes Simple Moving Average to alert developers when abnormal patterns appear. Our case study suggests that this technique is able to detect correctly a real problem in Facebook Message that annoyed many of its users. We conclude that it has the potential to reduce developer's effort in real life situation.
	
	Finally, from the evaluations and case studies, we suggest that using {\tool} Framework would help developers to map their concerns/interests with user's via the common expression of keywords, thus save them time and effort for discovering and understanding user's opinions.

\section*{Acknowledgment}
The authors would like to thank Dr. Young-Woo Kwon, Anand Ashok Bora, Sima Mehri Kotchaki, and Jiin Kim at Utah State University for their help in our evaluation. We also give our special thank to Dr. Tam Vu, Dr. Thang Dinh, and Ty Nguyen for their contributions to a prior study from which our work is inspired.






\bibliographystyle{IEEEtran}
\bibliography{MARK}

\begin{thebibliography}{10}
\providecommand{\url}[1]{#1}
\csname url@samestyle\endcsname
\providecommand{\newblock}{\relax}
\providecommand{\bibinfo}[2]{#2}
\providecommand{\BIBentrySTDinterwordspacing}{\spaceskip=0pt\relax}
\providecommand{\BIBentryALTinterwordstretchfactor}{4}
\providecommand{\BIBentryALTinterwordspacing}{\spaceskip=\fontdimen2\font plus
\BIBentryALTinterwordstretchfactor\fontdimen3\font minus
  \fontdimen4\font\relax}
\providecommand{\BIBforeignlanguage}[2]{{%
\expandafter\ifx\csname l@#1\endcsname\relax
\typeout{** WARNING: IEEEtran.bst: No hyphenation pattern has been}%
\typeout{** loaded for the language `#1'. Using the pattern for}%
\typeout{** the default language instead.}%
\else
\language=\csname l@#1\endcsname
\fi
#2}}
\providecommand{\BIBdecl}{\relax}
\BIBdecl

\bibitem{ARminer}
\BIBentryALTinterwordspacing
N.~Chen, J.~Lin, S.~C.~H. Hoi, X.~Xiao, and B.~Zhang, ``Ar-miner: Mining
  informative reviews for developers from mobile app marketplace,'' in
  \emph{Proceedings of the 36th International Conference on Software
  Engineering}, ser. ICSE 2014.\hskip 1em plus 0.5em minus 0.4em\relax New
  York, NY, USA: ACM, 2014, pp. 767--778. [Online]. Available:
  \url{http://doi.acm.org/10.1145/2568225.2568263}
\BIBentrySTDinterwordspacing

\bibitem{facebookStatistic}
\BIBentryALTinterwordspacing
``Number of monthly active facebook messenger users from april 2014 to june
  2015.'' [Online]. Available: \url{http://www.statista.com/
  statistics/417295/facebook-messenger-monthly-active-users/}
\BIBentrySTDinterwordspacing

\bibitem{word2vec}
\BIBentryALTinterwordspacing
T.~Mikolov, K.~Chen, G.~Corrado, and J.~Dean, ``Efficient estimation of word
  representations in vector space,'' \emph{CoRR}, vol. abs/1301.3781, 2013.
  [Online]. Available: \url{http://arxiv.org/abs/1301.3781}
\BIBentrySTDinterwordspacing

\bibitem{kmean}
J.~MacQueen \emph{et~al.}, ``Some methods for classification and analysis of
  multivariate observations,'' in \emph{Proceedings of the fifth Berkeley
  symposium on mathematical statistics and probability}, vol.~1, no.~14.\hskip
  1em plus 0.5em minus 0.4em\relax Oakland, CA, USA., 1967, pp. 281--297.

\bibitem{manning2008introductionInfoRetrieval}
\BIBentryALTinterwordspacing
C.~D. Manning, P.~Raghavan, and H.~Sch{\"u}tze, \emph{Introduction to
  information retrieval}.\hskip 1em plus 0.5em minus 0.4em\relax Cambridge
  university press Cambridge, 2008, vol.~1. [Online]. Available:
  \url{http://nlp.stanford.edu/IR-book/pdf/irbookonlinereading.pdf}
\BIBentrySTDinterwordspacing

\bibitem{Wiscom}
\BIBentryALTinterwordspacing
B.~Fu, J.~Lin, L.~Li, C.~Faloutsos, J.~Hong, and N.~Sadeh, ``Why people hate
  your app: Making sense of user feedback in a mobile app store,'' in
  \emph{Proceedings of the 19th ACM SIGKDD International Conference on
  Knowledge Discovery and Data Mining}, ser. KDD '13.\hskip 1em plus 0.5em
  minus 0.4em\relax New York, NY, USA: ACM, 2013, pp. 1276--1284. [Online].
  Available: \url{http://doi.acm.org/10.1145/2487575.2488202}
\BIBentrySTDinterwordspacing

\bibitem{facebook}
\BIBentryALTinterwordspacing
``Facebook messenger battery drain! - facebook's developer confirmed,'' Feb 12,
  2015. [Online]. Available: \url{http:// android forums. com/ threads/
  facebook- messenger- battery- drain. 902687/}
\BIBentrySTDinterwordspacing

\bibitem{faroo}
\BIBentryALTinterwordspacing
``Faroo spell corrector.'' [Online]. Available: \url{http://blog. faroo.
  com/2012/ 06/07/improved-edit-distance-based-spelling-correction/}
\BIBentrySTDinterwordspacing

\bibitem{norvicspell}
\BIBentryALTinterwordspacing
``Peter norvic spell corrector.'' [Online]. Available: \url{http:// norvig.
  com/ spell -correct .html}
\BIBentrySTDinterwordspacing

\bibitem{porter2001snowball}
M.~Porter and R.~Boulton, ``Snowball stemmer,'' 2001.

\bibitem{manning:SFlemma}
\BIBentryALTinterwordspacing
C.~D. Manning, M.~Surdeanu, J.~Bauer, J.~Finkel, S.~J. Bethard, and
  D.~McClosky, ``The {Stanford} {CoreNLP} natural language processing
  toolkit,'' in \emph{Proceedings of 52nd Annual Meeting of the Association for
  Computational Linguistics: System Demonstrations}, 2014, pp. 55--60.
  [Online]. Available: \url{http://www.aclweb.org/anthology/P/P14/P14-5010}
\BIBentrySTDinterwordspacing

\bibitem{SFpostagger}
\BIBentryALTinterwordspacing
K.~Toutanova, D.~Klein, C.~D. Manning, and Y.~Singer, ``Feature-rich
  part-of-speech tagging with a cyclic dependency network,'' in
  \emph{Proceedings of the 2003 Conference of the North American Chapter of the
  Association for Computational Linguistics on Human Language Technology -
  Volume 1}, ser. NAACL '03.\hskip 1em plus 0.5em minus 0.4em\relax
  Stroudsburg, PA, USA: Association for Computational Linguistics, 2003, pp.
  173--180. [Online]. Available:
  \url{http://dx.doi.org/10.3115/1073445.1073478}
\BIBentrySTDinterwordspacing

\bibitem{IrregularEnglishverbs}
\BIBentryALTinterwordspacing
``Irregular english verbs.'' [Online]. Available: \url{http:// www. us ing
  english. com/ reference/ irregular- verbs/}
\BIBentrySTDinterwordspacing

\bibitem{wiki1bil}
\BIBentryALTinterwordspacing
M.~Mahoney, ``English wikipedia dump.'' [Online]. Available:
  \url{http://mattmahoney.net/dc/textdata}
\BIBentrySTDinterwordspacing

\bibitem{miller1998wordnet}
G.~Miller and C.~Fellbaum, ``Wordnet: An electronic lexical database,'' 1998.

\bibitem{cocstatistic}
\BIBentryALTinterwordspacing
``Clash of clans daily revenue at 5.15 millions usd- hacker,'' February 11,
  2014. [Online]. Available: \url{http:// www. games industry. biz/ articles/
  2014-02-11-clash-of-clans-daily-revenue-at-5.15-million-hacker}
\BIBentrySTDinterwordspacing

\bibitem{gplayCrawler}
\BIBentryALTinterwordspacing
Akdeniz, ``An opensource googleplay crawler.'' [Online]. Available:
  \url{https:// github. com/ Akdeniz/ google-play-crawler}
\BIBentrySTDinterwordspacing

\bibitem{Vasa:preliminary}
\BIBentryALTinterwordspacing
R.~Vasa, L.~Hoon, K.~Mouzakis, and A.~Noguchi, ``A preliminary analysis of
  mobile app user reviews,'' in \emph{Proceedings of the 24th Australian
  Computer-Human Interaction Conference}, ser. OzCHI '12.\hskip 1em plus 0.5em
  minus 0.4em\relax New York, NY, USA: ACM, 2012, pp. 241--244. [Online].
  Available: \url{http://doi.acm.org/10.1145/2414536.2414577}
\BIBentrySTDinterwordspacing

\bibitem{hoon2013analysisReviewLandscape}
L.~Hoon, R.~Vasa, J.-G. Schneider, and J.~Grundy, ``An analysis of the mobile
  app review landscape: Trends and implications,'' Tech. rep., Faculty of
  Information and Communication Technologies, Swinburne University of
  Technology, Melbourne, Australia, Tech. Rep., 2013.

\bibitem{feedbackEmpirical}
D.~Pagano and W.~Maalej, ``User feedback in the appstore: An empirical study,''
  in \emph{Requirements Engineering Conference (RE), 2013 21st IEEE
  International}, July 2013, pp. 125--134.

\bibitem{Khalid:identifyIOS}
H.~Khalid, E.~Shihab, M.~Nagappan, and A.~Hassan, ``What do mobile app users
  complain about? a study on free ios apps,'' 2014.

\bibitem{iacob2013:complain}
C.~Iacob, V.~Veerappa, and R.~Harrison, ``What are you complaining about?: a
  study of online reviews of mobile applications,'' in \emph{Proceedings of the
  27th International BCS Human Computer Interaction Conference}.\hskip 1em plus
  0.5em minus 0.4em\relax British Computer Society, 2013, p.~29.

\bibitem{apiVSRating}
G.~Bavota, M.~Linares-Vasquez, C.~Bernal-Cardenas, M.~Di~Penta, R.~Oliveto, and
  D.~Poshyvanyk, ``The impact of api change- and fault-proneness on the user
  ratings of android apps,'' \emph{Software Engineering, IEEE Transactions on},
  vol.~41, no.~4, pp. 384--407, April 2015.

\bibitem{martin:appsamplingprob}
W.~Martin, M.~Harman, Y.~Jia, F.~Sarro, and Y.~Zhang, ``The app sampling
  problem for app store mining.''

\bibitem{Chandy:Spam}
\BIBentryALTinterwordspacing
R.~Chandy and H.~Gu, ``Identifying spam in the ios app store,'' in
  \emph{Proceedings of the 2Nd Joint WICOW/AIRWeb Workshop on Web Quality},
  ser. WebQuality '12.\hskip 1em plus 0.5em minus 0.4em\relax New York, NY,
  USA: ACM, 2012, pp. 56--59. [Online]. Available:
  \url{http://doi.acm.org/10.1145/2184305.2184317}
\BIBentrySTDinterwordspacing

\bibitem{Carreno}
L.~Galvis~Carreno and K.~Winbladh, ``Analysis of user comments: An approach for
  software requirements evolution,'' in \emph{Software Engineering (ICSE), 2013
  35th International Conference on}, May 2013, pp. 582--591.

\bibitem{guzman2014:reviewsAnalysis}
E.~Guzman and W.~Maalej, ``How do users like this feature? a fine grained
  sentiment analysis of app reviews,'' in \emph{Requirements Engineering
  Conference (RE), 2014 IEEE 22nd International}.\hskip 1em plus 0.5em minus
  0.4em\relax IEEE, 2014, pp. 153--162.

\bibitem{LDA}
\BIBentryALTinterwordspacing
D.~M. Blei, A.~Y. Ng, and M.~I. Jordan, ``Latent dirichlet allocation,''
  \emph{J. Mach. Learn. Res.}, vol.~3, pp. 993--1022, Mar. 2003. [Online].
  Available: \url{http://dl.acm.org/citation.cfm?id=944919.944937}
\BIBentrySTDinterwordspacing

\bibitem{SentiStrength}
M.~Thelwall, K.~Buckley, G.~Paltoglou, D.~Cai, and A.~Kappas, ``Sentiment
  strength detection in short informal text,'' \emph{Journal of the American
  Society for Information Science and Technology}, vol.~61, no.~12, pp.
  2544--2558, 2010.

\bibitem{Iacob:mara}
\BIBentryALTinterwordspacing
C.~Iacob and R.~Harrison, ``Retrieving and analyzing mobile apps feature
  requests from online reviews,'' in \emph{Proceedings of the 10th Working
  Conference on Mining Software Repositories}, ser. MSR '13.\hskip 1em plus
  0.5em minus 0.4em\relax Piscataway, NJ, USA: IEEE Press, 2013, pp. 41--44.
  [Online]. Available: \url{http://dl.acm.org/citation.cfm?id=2487085.2487094}
\BIBentrySTDinterwordspacing

\bibitem{nigam2000text:EMNB}
K.~Nigam, A.~K. McCallum, S.~Thrun, and T.~Mitchell, ``Text classification from
  labeled and unlabeled documents using em,'' \emph{Machine learning}, vol.~39,
  no. 2-3, pp. 103--134, 2000.

\bibitem{buggyappchecker}
M.~Gomez, R.~Rouvoy, M.~Monperrus, and L.~Seinturier, ``A recommender system of
  buggy app checkers for app store moderators,'' Ph.D. dissertation, Inria
  Lille, 2014.

\end{thebibliography}
%
%
%

\end{document}